\documentclass[a4paper,11pt]{article}
\usepackage[scale={0.8,0.9},centering,includeheadfoot]{geometry}
\usepackage{authblk}
\usepackage[normalem]{ulem}

\title{Spatial modelling with R-INLA: A review}

\author[1]{Haakon Bakka}
\author[1]{H{\aa}vard Rue}
\author[2]{Geir-Arne Fuglstad}
\author[2]{Andrea Riebler}
\author[3]{David Bolin}
\author[4]{Janine Illian}
\author[5]{Elias Krainski}
\author[6]{Daniel Simpson}
\author[7]{Finn Lindgren}
\affil[1]{CEMSE Division,
    King Abdullah University of Science and Technology,
    Saudi Arabia}
\affil[2]{Department of Mathematical Sciences,
    Norwegian University of Science and Technology,
    Norway}
\affil[3]{Department of Mathematical Sciences,
    Chalmers University of Technology,
    Sweden}
\affil[4]{CREEM, School of Mathematics and Statistics, 
        University of St Andrews, UK}
\affil[5]{Departamento de Estat\' istica,
    Universidade Federal do Paran\' a,
    Brazil}
\affil[6]{Department of Statistical Sciences,
    University of Toronto,
    Canada}
\affil[7]{School of Mathematics,
    University of Edinburgh,
    Scotland}

\usepackage{natbib}
\usepackage{amsthm} 
\usepackage{amsfonts} 
\usepackage{amscd} 
\usepackage{amsmath} 
\usepackage{graphicx}
\usepackage{epstopdf}
\usepackage{pgf}
\usepackage{array, multirow}
\usepackage{caption}
\usepackage{subcaption}

\usepackage{hyperref}

\usepackage[utf8]{inputenc}

\theoremstyle{definition}

\newtheorem{exmp}{Example}[section]

\newcommand{\m}[1]{\ensuremath{\mathcal{#1}}}
\newcommand{\mb}[1]{\ensuremath{\mathbb{#1}}}

\renewcommand{\d}{\ \mathrm{d} }

\newcommand{\up}[1]{\text{#1}}

\newcommand{\mat}[1]{\text{\bf #1}}

\renewcommand{\vec}[1]{\ensuremath{\boldsymbol{ #1}}}

\newcommand{\stencil}{\propto}
\newcommand{\neig}{\sim}

\DeclareMathOperator{\diag}{diag}

\graphicspath{{fig/}}

\def\red#1{#1}
\def\db#1{#1}

\begin{document}

\maketitle

\begin{abstract}
    Coming up with Bayesian models for spatial data is easy, but
    performing inference with them can be challenging. Writing fast
    inference code for a complex spatial model with
    realistically-sized datasets from scratch is time-consuming, and
    if changes are made to the model, there is little guarantee that
    the code performs well. The key advantages of R-INLA are the ease
    with which complex models can be created and modified, without the
    need to write complex code, and the speed at which inference can
    be done even for spatial problems with hundreds of thousands of
    observations.

    R-INLA handles latent Gaussian models, where fixed effects,
    structured and unstructured Gaussian random effects are combined
    linearly in a linear predictor, and the elements of the linear
    predictor are observed through one or more likelihoods. The
    structured random effects can be both standard areal model such as
    the Besag and the BYM models, and geostatistical models from a
    subset of the Mat\'ern Gaussian random fields. In this review, we
    discuss the large success of spatial modelling with R-INLA and the
    types of spatial models that can be fitted, we give an overview of
    recent developments for areal models, and we give an overview of
    the stochastic partial differential equation (SPDE) approach and
    some of the ways it can be extended beyond the assumptions of
    isotropy and separability. In particular, we describe how slight
    changes to the SPDE approach leads to straight-forward approaches
    for non-stationary spatial models and non-separable space-time
    models.
\end{abstract}

\section{Introduction}

Spatial modelling is an important, but computationally challenging,
statistical field. The main challenge is that the most common
modelling tool for capturing spatial dependency, the Gaussian random
field (GRF), is hard to use when there is a lot of data. A number of
strategies have been proposed for solving this problem, see
\citet{heaton2017methods} for an up-to-date review. In this paper, we
review one particular suite of methods, known as the \emph{SPDE
    approach} (stochastic partial differential equations approach). 
This approach, based on some advanced tools from the
theory of stochastic processes, has a computationally efficient
implementation in the R-package \texttt{INLA} (R-INLA)
\citep{art451,art632} and has been widely used in practice. The
computational efficiency of the R-INLA implementation, as well as the
relative simplicity of the interface, has allowed applied spatial
researchers to fit a broad range of spatial models to a wide array of
applications.

A GRF is completely determined through its mean and covariance
(-matrix or -function), and the theory is well understood.
Computationally, inference with GRFs naturally results in vector and
matrix algebra, for which we can use standard computer libraries. In
statistical modelling, we typically use a GRF with a parametrised
covariance structure; often a subset of the Mat\'ern covariance
family. In this paper, this set of parameters is referred to as
\emph{hyper-parameters}, and the parametrised GRF is referred to as a
\emph{spatial model component} or a \emph{spatial random effect}.

There are several ways to do inference using the covariance structure
of a continuously indexed GRF. The traditional way is to construct the
covariance matrix $\boldsymbol \Sigma $ for the GRF, based on the
observation locations, directly from the covariance function, and then
combine it with the covariance matrices for the other model components
to create a full covariance matrix for the observations, or for the
underlying latent model. This approach can work well for inference
with hundreds of locations, but for inference with a hundred thousand
locations, the approach is not computationally feasible since
computations with $\boldsymbol \Sigma$ are too time consuming. Beyond
the computational issues, it is challenging to create covariance
functions for other geometries such as spheres (the surface of the
earth), to introduce non-stationarity in the covariance function, and
to extend a spatial covariance function to a non-separable space-time
structure.

R-INLA turns the focus from the covariance matrix to the precision
matrix $\mat Q = \boldsymbol \Sigma^{-1}$, as it can be shown that
several common models with complex covariance structures have sparse
precision matrices \citep{book80}. Sparse matrices have
mostly zeroes, and (most of) the zeroes are not stored in the
computer. 
The sparsity structure of the precision matrix relates to conditional 
independence between the random variables in the multivariate 
Gaussian distribution \citep{book80}.
As we will show in this paper, it is possible to formulate
both discretely and continuously indexed spatial models with sparse
precision matrices. Let the dimension of a precision matrix for a
two-dimensional spatial model with spatial sparsity structure be
$n \times n$. Drawing samples, computing the normalising constant, or
doing inference, using the sparse precision matrix then require
computations of order $\m O(n^{3/2})$, compared to the more expensive
$\m O(n^2) $ storage and $\m O(n^3) $ computations required for the
corresponding dense covariance matrix. Applications with hundreds of
thousand of locations are then feasible, and applications with a few
thousand observations can be analysed with the click of a button.

In R-INLA we achieve the desired spatial sparsity structure for the
precision matrix of the continuously indexed GRF by using the
SPDE approach
\citep{lindgren2011explicit}. Instead of constructing a discrete model
for the GRF on a set of locations or grid cells by using a covariance
function, we construct a continuously indexed approximation of the GRF
by using a continuous model---an SPDE---that is defined on the entire
study area. For example, the Mat\'ern covariance function is given by
\begin{equation}
        c(s_1, s_2) = \sigma^2\frac{2^{1-\nu}}{\Gamma(\nu)}(\sqrt{8\nu}||s_1-s_2||/\rho)K_\nu(\sqrt{8\nu}||s_1-s_2||/\rho),
        \label{eq:matern-cov}
\end{equation}
where $\Gamma$ is the Gamma function, $\rho$ is the spatial distance
at which correlation is approximately $0.13$, $\sigma$ is the marginal
standard deviation, $\nu$ is the smoothness parameter, and $K_\nu$ is
the modified Bessel function of the second kind, order $\nu$. The
parameters of this covariance function have clear physical meanings,
but the covariance function can only be used directly for small
problems.

We use the result by \citet{whittle1954stationary,art455} that shows
that a stationary solution of the SPDE
\begin{align}
  \left(\kappa^ 2 - \Delta \right)^ {\alpha/2} (\tau u(\vec s)) = \m W (\vec s), \quad s\in\mb R^d,
  \label{eq:matern-spde}
\end{align}
\red{
has a Mat\'ern covariance function, where $\kappa>0, \tau>0$,
$\alpha > d/2$, $\Delta = \sum_i \partial^2/\partial s_i^2$ is the
Laplacian, and $\m W$ is standard Gaussian white noise. 
The parameters used in Equation \ref{eq:matern-spde} are the standard parameters for the SPDE and are
different from the ones in Equation \eqref{eq:matern-cov}, but there is a
one-to-one correspondence between them. The advantage of the parameters of the 
Mat\'ern covariance function is that they have direct physical interpretations and 
the advantage of the parameters of the SPDE is that they make it simpler to
write the SPDE.
See \citet{lindgren2011explicit} for the formulas describing how to convert between the parametrisations.}
Based on this
SPDE, restricted to a bounded domain with appropriate boundary
conditions, we can construct a continuously indexed approximation to
the solution that approximately has the Mat\'ern covariance structure.
We can also use the one-to-one correspondence between the parameters
of the Mat\'ern covariance function and the parameters of the SPDE, to
estimate the model using the computationally efficient approximation,
and interpret the results through the well-known Mat\'ern covariance
function. The use of a finite element method (FEM) for constructing
the approximate solution on the bounded domain allows for boundaries
made up of complex polygons and for different fidelity of the
discretisation in different areas. For integer $\alpha$ values, the
continuous domain SPDE solutions are Markovian, which is reflected in
sparse precision matrices for the discrete approximations. For other
$\alpha$ values, sparse approximations can yield close
correspondences.

We highlight the main advantages of the SPDE approach.
\begin{enumerate}
\item The dimension of the finite-dimensional Gaussian approximation
    to the solution of the SPDE only depends on the desired resolution
    and is invariant to the number of observations.
\item The non-vanishing spatial correlations of the approximate
    solution are represented by a Markovian structure on the precision
    matrix (inverse covariance matrix) where only close neighbours are
    non-zero. We present this idea in detail in Section
    \ref{sect-continuous-models}.
\item The non-zero structure of the precision matrix is invariant to
    the spatial correlation range, determined by $\kappa$, of the
    approximate solution. The \red{number} of non-zero neighbours does
    however depend on the smoothness $\nu$.
\item Small changes to the differential operator in the SPDE leads to
    models on manifolds, and non-stationary and non-separable models,
    see Section \ref{sect-advanced}. If the operator is changed, but
    the exponent $\alpha/2$ is the same, the process of computing the
    matrices is the same as for the stationary models, and the
    precision matrices are automatically sparse and positive definite
    as long as the SPDE is well-defined.
\end{enumerate}
The speed-up we get by running our spatial models in R-INLA, the ease
of using the spatial model together with other model components, and
the ability to use a wide variety of observation likelihoods for the
latent process makes R-INLA a very useful tool for applied statistical
modelling. Models that used to be too complex to be fitted in the
Bayesian framework are now possible to run in a day. And, perhaps more
importantly, models we could previously run in a day can now be run in
an even shorter time, enabling researchers to fit several different
models, to understand the data, to investigate prior sensitivity, to
investigate sensitivity to the choice of observation likelihood, to
run bootstrap analyses, and to perform cross-validation and other
predictive comparisons. Another improvement is in reproducibility, and
code checking by other researchers, as code published together with
papers can often be run in an hour.

There are several high impact applications using spatial models in
R-INLA; in the journal \emph{The Lancet} \citet{art615} performed a
space-time analysis of transmission intensity of malaria and
\citet{golding2017mapping} modeled under-five mortality and neo-natal
mortality in multiple countries with separable space-time models for
different age groups; in the journal \emph{Science} \citet{art616}
studied the effects of fragmentation on infectious disease dynamics;
in the journal \emph{Nature} \citet{art618} analysed the effect of the
various efforts to to control malaria in Africa. R-INLA's spatial
capabilities were also a key tool used by \citet{shaddick2018data} to
produce the global estimates of ambient exposure to ultra-fine
particulate matter of less than 2.5$\mu$m in diameter, known as
PM$_{2.5}$, that were used in both the 2016 Global Burden of Disease
study \citep{gakidou2017global} and the World Health Organization's
assessment of health risk due to ambient air pollution
\citep{world2016ambient}.

\red{
A collection of some recent examples of spatial applications with the R-INLA software, 
intended as a source of inspiration for the reader, follows;}
environmental risk factors to liver Fluke in cattle
\citep{innocent2017combining} 
using a spatial random effect to account
for regional residual effects; 
modelling fish populations that are recovering
\citep{boudreau2017connectivity} with a separable space-time
model; 
mapping gender-disaggregated development indicators
\citep{bosco2017exploring} using a spatial model for the residual
structure;
environmental mapping of soil \citep{huang2017evaluating} comparing a
spatial model in R-INLA with ``REML-LMM'';
changes in fish distributions \citep{thorson2017relative};
febrile illness in children \citep{dalrymple2017quantifying};
dengue disease in Malaysia \citep{naeeim2017estimating}; modelling
pancreatic cancer mortality in Spain using a spatial
gender-age-period-cohort model \citep{etxeberria-etal-2017}; soil
properties in forest \citep{beguin2017predicting} comparing spatial
and non-spatial approaches;
ethanol and gasoline pricing \citep{laurini2017spatio} using a
separable space-time model;
fish diversity \citep{fonseca2017identifying} using a spatial GRF to account for unmeasured covariates; 
a spatial model of unemployment \citep{pereira2017unemployment};
distance sampling of blue whales \citep{yuan2017point} using a
likelihood for point processes;
settlement patterns and reproductive success of prey
\citep{morosinotto2017competitors}; 
cortical surface fMRI data \citep{mejia2017bayesian} computing
probabilistic activation regions;
distribution and drivers of bird species richness
\citep{dyer2017global} with a global model, and comparing several
different likelihoods;
socio-environmental factors in influenza-like illness
\citep{lee2017socio};
global distributions of Lygodium microphyllum under projected climate
warming \citep{humphreys2017bayesian} using a spatial model on the
globe; 
logging and hunting impacts on large animals
\citep{roopsind2017logging};
socio demographic and geographic impact of HPV vaccination
\citep{rutten2017population}; a combined analysis of point and area
level data \citep{moraga-etal-2017};
probabilistic prediction of wind power
\citep{lenzi2017spatial}; 
animal tuberculosis \citep{gortazar2017animal};
polio-virus eradication in Pakistan \citep{mercer2017spatial} with a
Poisson hurdle model;
detecting local overfishing \citep{carson2017local} from the posterior
spatial effect;
joint modelling of presence-absence and abundance of hake
\cite{paradinas2017spatio}; 
topsoil metals and cancer mortality \citep{lopez2017compositional}
with spatially misaligned data;
\red{
applications in spatial econometrics \citep{art520,art584,col33}; 
modeling landslides as point processes \citep{lombardo2017point};
comparing avian influenza virus in Vietnamese live bird markets
\citep{mellor2018comparative} and predicting extreme rainfall events in
space and time \citep{art642}.}

This paper is meant to give an understanding of the possibilities and
limitations for spatial models in R-INLA. The main focus is on the
continuously indexed spatial models defined through the SPDE approach,
but we provide a brief description of areal models and recent
developments for them. We do not give a detailed introduction of
R-INLA, which is found in \citet{art632}, or how to program spatial
models in R-INLA, which is found in
\citet{lindgren2015bayesian,book125,art528,col33,
    krainski2017r} and at \url{www.r-inla.org}. After giving some
necessary background in Section \ref{sect-background}, we start with
areal models in Section \ref{sect-area-models} and proceed to discuss
continuously indexed models and the SPDE approach in Section
\ref{sect-continuous-models}. In Section \ref{sect-spatial-modelling}
we discuss how the spatial random effects together with general
features of R-INLA makes it possible to create a wide variety of
models ranging from simple Gaussian geostatistical models to spatial
point process models. In Section \ref{sect-advanced} we show how we
can loosen the restrictions of isotropy and Gaussianity for the
spatial effect within the SPDE approach, and we end with a discussion
on the road into the future in Section
\ref{sect-discussion}. 

\section{Notation and background on R-INLA}
\label{sect-background}

In this section we give a brief overview of the parts central to
spatial models from the R-INLA review paper \citep{art632}. We use
boldface symbols ($\vec u$) to denote the discrete (matrix or vector)
objects, while indexed lowercase ($u_i$) denotes the elements in the
matrices and vectors, and the ordinary lowercase ($u$) denotes a
continuous spatial field ($u(\vec s)$) or a function.

The data $y_i$ are conditionally independent given the (linear)
predictor $\eta_i$
$$ y_i | \eta_i, \vec \theta_0 \sim \pi(y_i | \eta_i, \vec \theta_0),$$
where different $y_i$ can have different observation likelihoods.
The vector $\vec \theta_0$ is the first set of hyper-parameters, and are usually the scale and shape parameters of the chosen likelihood(s).

The linear predictor $\vec \eta$ is modelled by a sum of fixed and
random effects
$$\vec \eta = \mat A_1 \vec u_1 + ... + \mat A_k \vec u_k + ... $$
where $k$ signifies that this is component number $k$, but we will
suppress the $k$ for notational convenience. The random vector $u$ has
a Gaussian prior
$$\vec u | \vec \theta_k \sim \m N(\mat 0, \mat Q^{-1}) $$
where the precision matrix $\mat Q$ depends on the hyper-parameters
$\vec \theta_k$ for this random effect. There must be a constant
sparsity structure for $\mat Q$, across all of $\vec \theta_k$'s
values, and we call this the graph (the non-sparse elements in
$\mat Q$ can be zero for some values of $\vec \theta_k$). The
projection matrix $\mat A$ is a known sparse matrix, often just a
matrix of 0's and 1's signifying which entry of the random effect
$\vec u$ is used for the observation $i$. For example, if both the
first and seventh observation in space is at the same location, the
first and seventh row of $\mat A$ are equal and have a 1 in the column
corresponding to that location and 0's otherwise.
We detail several different choices of models for $u$ in this review, 
and we describe the $\mat A$-matrix for SPDE models in \red{Section} \ref{sect-understanding-prec}.

The dimension of the hyper-parameters
$$\vec \theta = (\vec \theta_0, \vec \theta_1, ..., \vec \theta_K)$$
should not be too large, \red{meaning less than 20 and preferably around 5},
\red{because the exploration of the posterior of $\vec \theta$ is expensive.
One evaluation of $\pi(\vec \theta | \vec y)$ requires factorising the precision matrix and 
approximating the posterior contribution of the likelihood by a Laplace approximation.}

Additionally, the posterior of the hyper-parameters should be unimodal
and not too different from a multivariate Gaussian. To satisfy this,
we depend on good parametrisations. For example, the posterior for the
marginal standard deviation $\sigma$ is usually skewed and has a heavy
tail, while the posterior of $\log(\sigma)$ is well-behaved.

In this framework there is \emph{no} difference between 1-dimensional
models, e.g.\ non-linear covariate effects, and 2-dimensional spatial
models, or 3-dimensional space-time models. The \texttt{inla()}-call
itself does not know that we are fitting a spatial model, it only
knows that we have a precision matrix with a certain graph and that
the supplied $\mat A$-matrix connects the precision matrix to
$\eta_i$. \emph{We} know that this precision matrix represents a
spatial correlation structure and that the supplied $\mat A$-matrix
projects from the 1-dimensional vector $\vec s$ of spatial indices to
the two dimensional spatial (longitude, latitude) description. The
work we need to do, to create a spatial model in R-INLA, is
constructing a ``good'' precision matrix and $\mat A$-matrix,
according to our understanding of ``good''.

For applications with space-time data, the simplest interaction models
are the separable models, defined through Kronecker products,
$$\mat Q_\mathrm{spacetime} = \mat Q_\mathrm{time}  \otimes \mat Q_\mathrm{space}.$$
Kronecker models are implemented as a general feature in R-INLA, where
$\mat Q_\mathrm{space}$ can be any spatial model, including the model
in Section \ref{sect-area-models}, and $\mat Q_\mathrm{time}$ can be
selected from a small collection of temporal models, including random
walk of order 1 and 2, autoregressive of order 1, and iid models
(replicates). The ability to mix and match model components to create
the desired space-time model is of great interest, and means that
whenever we implement a spatial model in R-INLA, we get all these
space-time models with almost no additional work.

The standard approach for making predictions is to add ``fake data
rows'' containing the covariates/location we wish to predict, but with
\texttt{NA} in place of observed data. This can be computationally
inefficient for spatial models since it is computationally equivalent
to fitting extra observations. For spatial predictions where there may
be $10^5$ prediction locations this may lead to long computation times
\citep{huang2017evaluating} and while it can be reduced somewhat by
re-running the model multiple times with disjoint subsets of the
desired prediction locations \citep{poggio2016bayesian}, a better
approach is to use i.i.d.\ samples from the joint posterior.
\citet{fuglstad2018environmental} show that similar prediction results
can be obtained in 11 minutes using posterior samples as takes 24
hours with the standard approach with \texttt{NA} observations
\citep{huang2017evaluating}.

\subsection{\red{The Laplace approximation}}

\red{
The Laplace approximation is an essential component of INLA, 
allowing for fast computations across a wide range of likelihoods and link functions.
If the likelihood is Gaussian, i.e. $\vec y \sim \m N(\vec \eta, \vec \theta_0 )$, 
the unnormalised posterior density of $\pi(\vec \theta | \vec y)$ can be computed exactly,
\begin{align}
\pi(\vec \theta | \vec y) \propto \frac{\pi(\vec y| \vec \eta= \vec 0, \vec \theta) \pi(\vec \eta=0 | \vec \theta)}{\pi(\vec \eta = \vec 0 | \vec y, \vec \theta)} \pi(\vec \theta). \label{eq-laplace1}
\end{align}
For non-Gaussian likelihoods, we compute an approximation of this equation.
The Laplace approximation does not approximate the likelihood $\pi(\vec y | \vec \eta, \vec \theta)$, but
the conditional distribution $\pi(\vec \eta | \vec y, \vec \theta)$.
For example, for a Poisson likelihood, a Gaussian prior on $\eta_i$, and observing $y_i=1$, the $\pi( y_i | \eta_i, \vec \theta)$ cannot be approximated by a Gaussian, but $\pi(\eta_i |  y_i, \vec \theta)$ can.
The Laplace approximation is a quadratic approximation of the log-density around the posterior mode of $\vec \eta | \vec \theta$, which is found by iteration.
This posterior mode is subsituted for $\vec 0$ in equation \eqref{eq-laplace1} to compute $\pi(\vec \theta | \vec y)$.}

\red{
        The posterior for $\vec \eta$ and all other latent variables are computed by 
        numerical integration over 
        $\pi(\vec \theta | \vec y)$, using an additional Laplace approximation, 
        see \citet{art632} Section 3.2.
        The option \texttt{int.strategy = 'eb'} in R-INLA specifies that this integration 
        is only done with a single evaluation at the posterior mode of $\vec \theta$, 
        and is commonly used for very computationally intensive models.
        The option \texttt{int.strategy = 'ccd'} is the default for $\vec \theta$ of 
        dimension larger than 2, and allows for computing the posterior of $\eta$ 
        when the dimension of $\theta$ is large (but less than 20).
        The option \texttt{int.strategy = 'grid'} can be used to create a grid of 
        integration points over $\vec \theta$, and the user can further configure 
        this grid to gain greater accuracy when representing the posterior.
        However, grids are afflicted by the curse of dimensionality and appropriate only for low-dimensional spaces.}

\red{
When the model includes a spatial model component $u$, 
the posterior marginals for $u$ are computed by R-INLA.
This collection of marginals is rarely used directly, e.g.\ it is almost impossible to plot the entire collection.
From these marginals, R-INLA computes quantiles and the marginal standard deviation.
The spatially varying posterior marginal standard deviation is often used as a proxy for spatial uncertainty.
In applications with a non-linear link function, however, the standard deviation is not a good proxy for uncertainty, 
and one can instead use the upper and lower quantiles, or the inter-quartile range.}

\red{
        \section{Areal models and other discretely indexed models}
}
\label{sect-area-models}

\red{
In this review we separate the models into discretely and continuously indexed models.
A discretely indexed model (component) $u$ has a finite set of indices $i$, 
and it is not clear how to extend the model to other indices, 
i.e. other `locations', 'areas' or `covariate values'.}
Discretely indexed spatial models fit into the standard framework of
R-INLA and there are several good introductions towards their
implementation, see for example \citet{schrodle-held-2011,
        schrodle-held-2011b, schrodle-etal-2012, ugarte-etal-2014} or
\citet[Section 6.1-6.4]{book125}. One of the most well known discrete
spatial model components is what we refer to as the \emph{Besag} model
\citep{besag1991bayesian}---in honour of the statistician J.E.
Besag---commonly known as the ``CAR'' or ``iCAR'' model. We use the
name conditionally autoregressive (CAR) model for any multivariate
Gaussian model that is built up through conditional distributions, so,
in our terminology, the Besag model is a particular example of a CAR
model.

\subsection{The Besag model}
\label{sect-besag-mod}

\begin{figure}
        \centering
        \includegraphics[width=.4\linewidth]
        {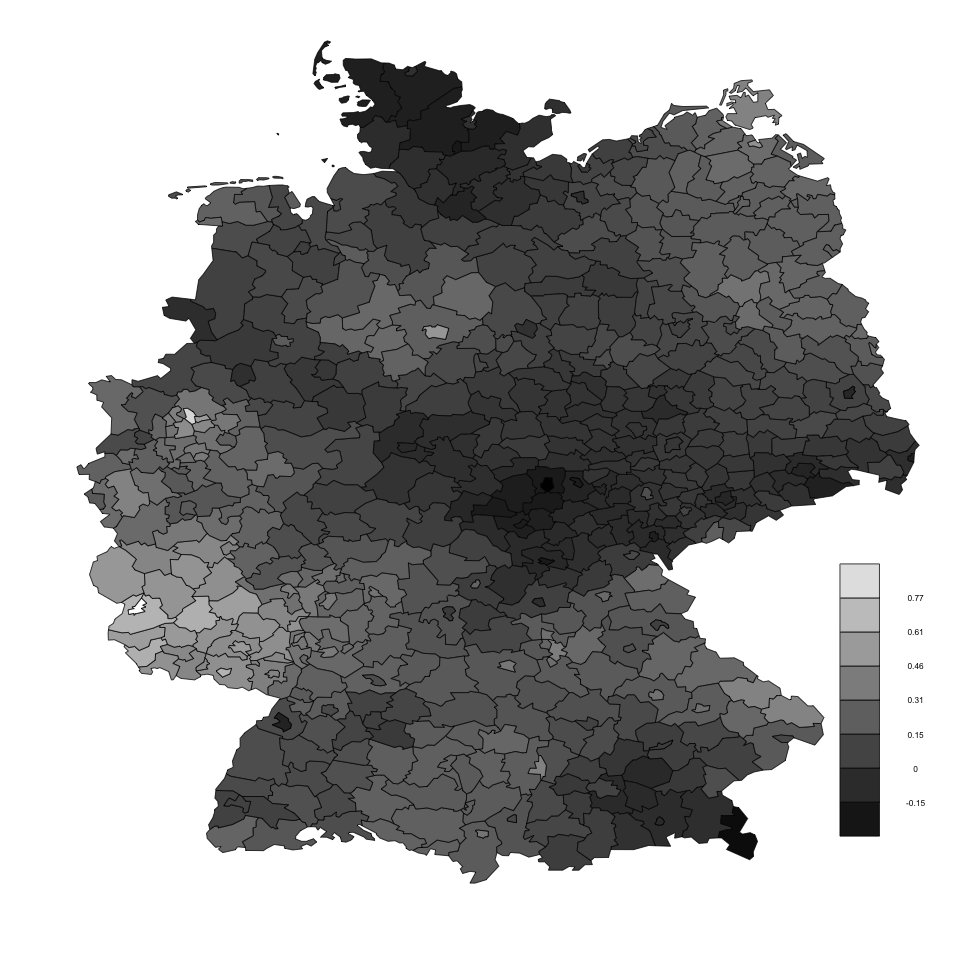}
        \includegraphics[width=.4\linewidth]{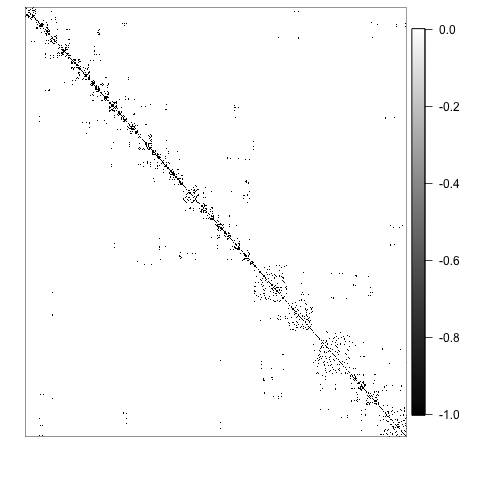}
        \caption{Besag model.
                The left plot shows an example dataset over regions in
                Germany, and the right plot shows the sparse precision
                matrix $\mat Q$.
        }
        \label{fig-besag}
\end{figure}

The Besag model models the values $u_i$ on a collection of
regions $i=1, \ldots, n$, see Figure \ref{fig-besag}, conditionally on
the neighbouring regions. Two regions are usually defined as
neighbours when the share a common border.
The conditional distribution for $u_i$ is
\begin{align*}
  u_i | \vec u_{-i}, \tau_u \sim \m N \left( \frac{1}{d_i} \sum_{j \neig
  i} u_j,
  \frac{1}{d_i} \frac{1}{\tau_u} \right),
\end{align*}
where $j \neig i$ denotes that $i$ and $j$ are neighbouring areas, and
$d_i$ is the number of neighbours. The joint distribution is given by
\begin{align*}
  \vec u | \tau_u\sim \m N \left( 0, \frac{1}{\tau_u} \mat Q^ {-1} \right),
\end{align*}
where $\mat Q$ denotes the structure matrix and is defined as 
\begin{align}
  Q_{i,j} = \left\lbrace 
  \begin{tabular}{ll}
    $d_i,$ & $i=j$ \\ 
    -1, & $i\neig j$ \\ 
  \end{tabular} \right. ,
  \label{eq:besagQ}
\end{align}
and $0$ otherwise. This structure matrix directly defines the
neighbourhood structure and is sparse per definition, having $\m O(n)$
non-zero elements. Of note, a weighted Besag model, where positive
weights are incorporated for each pair of neighbouring regions, can be
defined analogously, see \citet[Section 3.3.2]{book80}. There are
different ways to provide the neighbourhood structure given by
\eqref{eq:besagQ} to R-INLA: 
\red{
1) Through an ASCII file; 2) through a symmetric adjacency matrix of dimension
$n \times n$;}
3) by extracting the structure
directly from the shapefile using the R-packages {\tt maptools}
\citep{maptools} and {\tt spdep} \citep{bivand-etal-2015-2,
    bivand-etal-2013}. 
We refer to \citet[Section 6.1]{book125} for
implementation details of the Besag model in R-INLA.

It is generally recommended to combine the Besag model with an
additional unstructured random effect
$v_i\mid \tau_v \sim \mathcal{N}(0, \tau_v^{-1})$ per region
$i=1, \ldots, n$. The resulting spatial model $u_i + v_i$ for
$i=1, \ldots, n$ is often termed BYM model following the initials of
\red{
the authors who proposed it, 
Besag, J., York, J., and Mollie, A. \citep{besag1991bayesian}.}

\subsection{Prior specification for the Besag model}
\label{sect-besag-pri}

The Besag model penalises local deviation from a constant level in the
case where all regions are connected. The prior on the
hyper-parameter(s) will control this local deviation, i.e. the amount
of smoothing, but unfortunately its specification is not
straightforward, see for example \citet{bernardinelli-etal-1995,
    wakefield-2007}. One challenge is that $\tau_u$ is not directly
interpretable, as it depends on the underlying graph.

The first step to interpreting $\tau_u$ is making the Besag model
proper, by constraining the model to sum to zero (on each disconnected
set). 
\red{
The second step is to make the interpretation of $\tau_u$ independent of the
total number of regions and the set of edges (connections between}
neighbours). To see how unstable the definition of $\tau_u$ is
consider the marginal standard deviation
$$\up{Sd}(u_i) = \frac{\sqrt{{\mat Q}^{-}_{ii}}}{\sqrt{\tau_u}} \quad \text{for } i \in
1, \ldots, n, $$ whereby $\mat Q^{-}$ denotes the generalised inverse
of $\mat Q$. The marginal standard deviation is not constant over $i$,
and any maximum or average value depends heavily on the size $n$ of
the model. \citet{art521} propose to scale $\mat Q$ such that the
geometric mean of the marginal variances of $\mat u$ does not depend
on $n$ or on the connectivity structure. This implies that $\tau_u$
represents the precision of the (marginal) deviation from a constant
level, independently of the underlying graph, facilitating prior
specification. \citet{art645} recommend scaling each connected
component independently\red{, and give advice for how to define the 
Besag model for a general graph}.

\red{
Further, we can specify the joint prior for the Besag and iid components 
by reparametrising the BYM}
model as
\begin{equation}
  \mathbf{b} = \frac{1}{\sqrt{\tau}} \left(\sqrt{1-w} \vec{v} +
    \sqrt{w} \vec{u}^\star\right)
  \label{eq:bym2}
\end{equation}
containing a scaled Besag component $\vec{u}^\star$. This model,
called BYM2 in R-INLA, explicitly models the distribution of variance
between two components; the structured scaled effects $\vec{u}^\star$
and the unstructured effects $\vec v$. \red{For details on how to choose
sensible priors for $\tau$ and $w$ in the BYM2 model we refer to Appendix~\ref{app-besag}.}

\red{
\subsection{Discretely indexed models}
}

\red{
The Besag model is the most commonly used spatial discretely indexed model.
Any other discretely indexed model $u \sim \mathcal N(0, \tau^{-1} Q^{-1})$ is implemented as \texttt{generic0}.
Prior specification for this $\tau$ can be done in a similar manner to what we do for the Besag.
Discretely indexed models where the precision matrix depends on hyper-parameters can be implemented using \texttt{rgeneric}, such as the model proposed by
\citet{dean-etal-2001}, and the
Leroux model \citep{leroux-etal-2000}.
However, in these cases prior specification depends on the model in question, and we are often not able to give a recommendation.}

\red{
In the research done by the R-INLA group there is a general 
movement away from discretely indexed models, 
towards continuously indexed models. 
The discretely indexed models are typically
used because they offer a simple and computationally efficient way
to achieve spatial smoothing. 
However, these models ignore sub-regional variation,
do not account for differing region sizes or how much boundary is shared
between each pair of regions, 
and have difficulties handling geographic boundaries that are channging in time. 
To overcome these issues, one can model the data
at a very fine resolution using
continuously indexed models, 
and include the spatially aggregated observations
as integrals over the regions. 
However, this is challenging for non-linear link
functions, and when we model, for example, risk we require 
fine-scale information about the 
spatially varying population-at-risk.
Also, in some situations the areas are themselves relevant for 
modelling the underlying process (e.g. in \citet{lombardo2017point}).
Because of this, we expect that discretely indexed models will continue to be an important part of the R-INLA package going forwards.}

\subsection{Space-time models}

\red{
Extending spatial model components to space-time interaction components
is straightforward using
R-INLA, see Section \ref{sect-background}, and thus the 
Besag model can be extended to a space-time
interaction model.}
\citet{knorrHeld-2005} proposed four interaction types where the
structure matrix results as a Kronecker product of the structure
matrices of the effects supposed to interact. The resulting model can
\red{ 
be defined in R-INLA either as a user-defined \texttt{rgeneric} model}
or by making
use of the Kronecker structure, see \citet{riebler-etal-2012, art517}
for code snippets. The challenge of these interaction models is the
possibly large number of linear constraints required to ensure
identifiability of all model parameters \citep{schrodle-held-2011b,
    papoila-etal-2014, goicoa-etal-2017}. Since the computational cost
of constraints in R-INLA is $\m O(nk^ 2)$ for $k$ constraints and $n$
space-time points, a large number of constraints will result in models
that are no longer computationally feasible. \citet{elias2018}
proposes an alternative formulation avoiding linear constraints
entirely by trading them for computationally cheaper linear
combinations to solve identifiability issues. Future work aims at
defining those linear combinations automatically whenever an R-INLA
user uses one of the standard interaction models.

\section{Continuously indexed models}
\label{sect-continuous-models}

The Mat\'ern covariance function shown in Equation
\eqref{eq:matern-cov} is one of the most important and most frequently
used covariance functions for spatial models. 
\red{In this section} we
discuss how the Mat\'ern model is represented in R-INLA through the
SPDE approach.

\subsection{Representing a continuously indexed spatial model in
    R-INLA}
Let $u(\vec s)$ be a continuously indexed GRF and assume that
observations $y(\vec s_k)$ are made of a physical process that is
described by
\begin{equation}
        y(\vec s_k) =  X(\vec s_k) \beta + u(\vec s_k) + \epsilon_k,
        \label{eq:simple-geostat}
\end{equation}
where $X$ is a spatially varying covariate, $\beta$ is the coefficient
of the effect of the covariate, $\vec s_k$ is the observation
location, and $\epsilon_k$ is Gaussian noise that is i.i.d.~for the
observations. The first part of the model,
$X(\vec s_k)\beta + u(\vec s_k)$, is the spatially varying signal of
interest, but it cannot be directly observed due to the noise
$\epsilon_k$ associated with the observation process. More complex
observational processes can be constructed using a link function and a
non-Gaussian likelihood 
	as we explain in Section \ref{sect-spatial-modelling}.

The question is now how to represent the covariance structure of
$u(\vec s)$ in a computationally efficient way for performing
inference with the above model in R-INLA. For now, assume that the set
of observations form a square grid. The vector $\vec u$, which
consists of the variables stacked column-wise, has a multivariate
Gaussian distribution with a covariance matrix $\boldsymbol \Sigma$
computed from the covariance function. In R-INLA, it is represented as
$$\vec u  \sim \m N(\mat 0, \mat Q^{-1}), $$
where $\mat Q = \boldsymbol \Sigma^{-1}$.
However, the precision matrix $\mat Q$ is in general not sparse, 
which makes computations infeasible
for large datasets.

Conditional Autoregressive (CAR) and Simultaneous Autoregressive (SAR)
models \citep{art122} have sparse precision matrices by construction,
and a first attempt would be to approximate the desired covariance
structure for the gridded observations by such models. One way to find
a CAR to represent the GRF we want is to make a long list of different
spatial CAR models and investigate what covariance functions they
approximate. Along this line, \citet{art197} assumed that the
parameters in the precision matrix for grid cells more than $k$ steps
apart in at least one direction was zero, and parametrised the rest of
the precision matrix. Assuming translational and rotational invariance
(90 degree rotations), there are only a few parameters that need to be
inferred; for $k=3$ there are 6 parameters.

This approach to parametrise CAR models has several issues. Any
parametrisation of the CAR model must give positive definite precision
matrices, but the space of valid parameters does not have an intuitive
shape (see e.g.\ \citet{book80} Section 2.7.1). One needs to
map the parameters of the CAR model to interpretable parameters, such
as the spatial range parameter, in a continuous way, but this is
difficult and may have to be done separately for each application.
Further, it is necessary to investigate for which parts of the
parameter space the approximation to the continuous model is
sufficiently good, and, setting priors on the CAR parameters
	necessitates dealing
with the boundaries between
proper and intrinsic models. 
Perhaps most importantly, generalising to
irregular observations, to the sphere, or to non-stationary covariance
functions, is notoriously difficult as it exacerbates 
  all these issues.

The key to solving these issues is to stop focusing on the parameters
of the CAR model, and instead focus on the continuous representation
of the GRF
through an SPDE, letting the CAR parameters be a side-effect of
computing a continuous approximation to the continuously indexed GRF.
The SPDE approach produces precision matrices that enjoy the good
computational properties of the CAR models and is valid for any set of
observation locations. Generalisation, interpretable parameters, and
stability of the CAR structure, can then be investigated based on the
continuous interpretation of the SPDE.

\subsection{Discretising a differential operator}

In this section we consider functions on $\mb R^2$ and denote the
coordinates of locations $\vec s \in \mb R^2$ by $x$ and $y$. A
differential operator $\m L$ is a function of functions, taking as
input a surface $u(\vec s)$ and gives as output another surface
$\m L(u)(\vec s)$, for example
$$\m L_1(u) = \kappa^2 u - \nabla\cdot\nabla u, $$
where $\kappa>0$,
$\nabla = \left[\frac{\partial u}{\partial x}, \frac{\partial
      u}{\partial y}\right]$ is the gradient, and
$\Delta=\nabla\cdot\nabla=\partial^2/\partial x^2
+ \partial^2/\partial y^2$ is the Laplacian. Using $\m L_1$ on the
function $u(x,y) = \sin(2x)+y$ produces
$L(u)(x,y) = \kappa^ 2 \sin(2x) + \kappa^2 y + 4\sin(2x) $.

When $u$ is defined on a grid, we can approximate the continuously
indexed $u$ by a vector $\vec u$ consisting of the values of $u$ on
the grid stacked column-wise. For this discrete representation, the
operator $\m L$ has a corresponding discretised version that can be
written as a matrix $\mat L$, so that $\mat L \vec u$ produces
approximately the same values as if $\m L u$ were evaluated at the
grid. There are many ways to discretise operators, giving many
different possibilities for $\mat L$-matrices.
One discretisation approach is the use of finite differences on a regular grid,
allowing the CAR to be visualised by a computational stencil, see Appendix \ref{app-grid}.

\subsection{Constructing a continuously indexed approximation}

\red{
Discretised operators based on finite differences for SPDEs can be
used to construct models that fit within the R-INLA framework.
However, finite difference methods give discretely indexed approximations,
having several disadvantages,}
including being
difficult to extend to irregular locations, non-rectangular grids or
sparsely observed grids. 
Therefore, it is more natural to follow the
numerics literature on partial differential equations (PDEs) and
instead approximate a solution of the SPDE as a sum of finitely many
basis functions. 
We now introduce a more complex discretisation method
that will be used to construct the approximately Mat\'ern GRFs based
on the SPDE in Equation \eqref{eq:matern-spde}.

Let $\{\phi_j(\vec s)\}$ be a set of basis functions, then any
permissible function $u(\vec s)$ is on the form
\begin{align}
  u(\vec s) = \sum_{j=1}^{J} a_j \phi_j(\vec s), \label{eq-basis}
\end{align}
where $a_j$ are real-valued coefficients for the basis functions. The
idea is to discretise the differential operator $\m L$ to a matrix
$\mat L$ on the coefficients $a_j$, instead of with respect to a grid.
In this setting,
\begin{align}
  \m L \left(\sum_{j=1}^{J} a_j \phi_j(\vec s)\right) \approx \sum_{j=1}^{J} b_j \phi_j(\vec s), \label{eq-operator-on-coeff}
\end{align}
where $(b_1, \ldots, b_J) = \mat L(a_1, \ldots, a_J)$. This idea can
be made precise in terms of bases, operators and projections on
Hilbert spaces. One of the main advantages of using a basis of
continuous functions is that any solution is a continuous function,
hence the solution is defined everywhere and can be evaluated at any
desired location without interpolation techniques. 
This is what we call a ``continuously indexed approximation''.
The main difficulty
with this approach is to find a good basis
$\left\lbrace \phi_j \right\rbrace$, with a computationally efficient 
(sparse)
discretisation matrix $\mat L$.

\begin{figure}
        \centering
        \includegraphics[width=.8\linewidth]
        {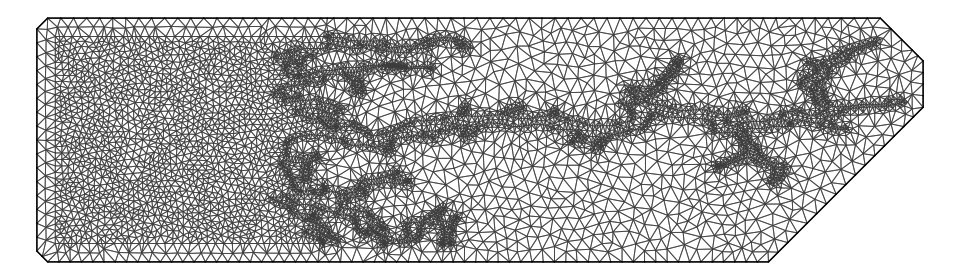}
        \caption{An example mesh, constructed for the Norwegian fjord ``Sognefjorden''.
        }
        \label{fig-mesh}
\end{figure}

The approach that was chosen in \citet{lindgren2011explicit} is known
as the finite element method (FEM) with linear elements, see e.g.\
\citet{brenner2007mathematical}, which is an excellent practical
computational tool. In the FEM, the grid is replaced by a \emph{mesh},
see Figure \ref{fig-mesh}. In our framework, the mesh is composed of
triangles and covers the entire domain, and a bit more to account for
boundary conditions. The basis functions, $\phi(\vec s)$ in equation
\eqref{eq-basis}, known as linear \emph{elements}, or hat-functions,
are constructed based on this mesh. The vertices of the triangles are
called nodes, and at each node $j$, $\phi_j=1$, at any other node
$\phi_j=0$, and $\phi_j$ on the triangles are given by linear
interpolation. Any piecewise linear---with respect to the
mesh---function is then a linear sum of these elements. 
The theory of the FEM
is well-established, meaning we can re-use results in the literature
to compute the matrix $\mat L$, and, given that we follow established
recommendations for how to define the mesh and the elements, we expect
no unwelcome surprises. Additionally, since the mesh is a collection
of triangles, we can have complicated polygons as boundaries to our
mesh.

\subsection{R-INLA implementation of the Mat\' ern fields}
\label{sect-cont-implement}

The construction of the mesh for a specific application is an involved
process, implemented as \texttt{inla.mesh.2d}, and documented in
R-INLA and in the tutorials at \url{www.r-inla.org}. The
$\mat A$-matrix connects the GRF-on-the-mesh to the GRF-on-the-data,
$A_{i,j}=\phi_j(s_i)$, and is computed using \\
\texttt{inla.spde.make.A(mesh, observation.locations)}. When smoothing
spatial data, with no covariates, the INLA-call is simple, the
$\mat A$-matrix is included in the \texttt{control.predictor} input.
However, if there are several other model components or a space-time
model component, keeping track of all the different $\mat A$-matrices
is challenging, and we developed the \texttt{inla.stack} interface to
handle all these matrix operations in a general way
\citep{lindgren2015bayesian}. In
Section~\ref{sect-cont-spatial-r-packages} we will mention some recent
efforts to further simplify the specification of these models.

After the mesh in constructed, the R-INLA model
\texttt{inla.spde2.pcmatern} implements the Mat\' ern GRFs, and is
described in \citet{lindgren2011explicit} Section 2.3, and
\citet{lindgren2015bayesian} Section 1.1. The stationary Mat\' ern
family is represented by an approximate solution of the SPDE
\begin{align}
  \label{eq-matern-general} \left(\kappa^ 2 + \nabla\cdot\nabla \right)^ {\alpha/2} (\tau u(\vec s))
  = \m W (\vec s), \qquad s \in \m D \subset \mb R^ d, 
\end{align}
where $\m D$ is a polygonal domain (i.e.\ we can make a mesh on
$\m D$), $\alpha$ is an integer, and we use Neumann boundary
conditions. The boundary conditions are not discussed in this paper,
but warrants that the mesh covers a larger area than the observations,
see \citet{lindgren2011explicit} \red{Appendix} A.4. This SPDE results in
smoothness $\nu = \alpha - d/2$. The basic FEM matrices are set up as
\begin{align}
  C_{i,i} &= \langle \phi_i, 1 \rangle\\
  G_{i,j} &= \langle \nabla \phi_i, \nabla \phi_j \rangle \\
  \mat K &= \kappa^ 2 \mat C + \mat G,
\end{align}
where the other elements of $\mat C$ are zero \citep{lindgren2011explicit}.
The SPDE is then discretised to
\begin{align}
  \mat Q_1 &= \mat K \\
  \mat Q_2 &= \mat K \mat C^ {-1} \mat K \\
  \mat Q_\alpha &= \mat K \mat C^ {-1} \mat Q_{\alpha-2} \mat C^ {-1} \mat K.
\end{align}
For $\alpha=2$, the default value in R-INLA, this can be written as
\begin{align}
  \mat Q = \tau^ 2 \left( \kappa^ 4 \mat C + 2\kappa^2 \mat G_1 + \mat G_2  \right), \label{eq-prec-a-2}
\end{align}
and the discretisation of $\m L_1$ on the linear FEM basis is
$\mat K$. 
We consider the approximations to be Markov because the precision matrices are sparse.
More details on the the finite element method, the weak formulation of the SPDE, 
and how to compute the different matrices can be found in \cite{bakka2018solve}.
The current implementation provides access to models in the
entire interval $\alpha\in(0,2]$, through the parsimonious fractional
approximation introduced in the Authors' Response to the discussion of
\citet{lindgren2011explicit}, which on 2-dimensional domains includes
the non-Markovian exponential covariance model, for $\alpha=3/2$. 

The discrepancy between the continuous domain SPDE solutions and the
finite basis approximations is a function of the mesh quality, both in
terms of the size and shape of the triangles. Generally speaking,
small and regularly shaped triangles give the smaller error, but the
discretisation error depends on the correlation scale and smoothness
of the random field. The exploratory tool \texttt{meshbuilder} in
R-INLA can be used to interactively construct meshes, and to assess
their discrete approximation properties. 

\subsection{Understanding the precision matrix}
\label{sect-understanding-prec}

\red{
Using R-INLA to compute the precision matrix in equation \eqref{eq-prec-a-2} 
is straightforward, 
but understanding what it is the precision matrix of, 
and how to extend it to new locations, less so.
This is the precision matrix of the coefficients of the elements, meaning that, 
to simulate from the GRF we first sample these coefficients
$$\tilde{\vec a} \sim \m N(\boldsymbol 0, \mat Q^{-1}) $$
and then we multiply these coefficients with the elements, as in equation 
\eqref{eq-basis}.
Alternatively, we assign to each mesh node $i$ the 
value $\tilde a_i$, and do a linear interpolation 
between the mesh nodes.
This gives a continuous function sampled from the approximate Mat\' ern field.
To construct spatial maps, we usually evaluate this function at a very fine grid.
}

\red{
For extending the model to
locations other than the mesh points, e.g.\ to fit the data,
one must first compute the $\mat A$-matrix, projecting 
from the mesh to these locations.
Each location has a corresponding row in the $\mat A$-matrix, and
any location that is exactly at a mesh node $j$ has a 0-row with a 1 in entry $j$.
A generic locations is always in one of the mesh triangles, 
and its matrix row can have up to three non-zero values,
which are found in
the three entries corresponding to 
the three mesh nodes at the corners of this triangle, 
see \citet{krainski2017r} for a detailed explanation.
Because $u(s)$ is a Gaussian field, 
any set of locations $ S = (s_k)_k $ gives a multivariate Gaussian vector $u(S)$.
The SPDE approximation of this vector has covariance matrix
$$ \boldsymbol \Sigma = \boldsymbol A \boldsymbol  Q^{-1} \boldsymbol A^\top. $$
}

\subsection{Areal observations}

A major advantage of the GRF in the SPDE approach with linear
elements, compared to the exact Mat\'ern model, is that integrals of
the GRF over an arbitrary area can be described as linear combinations
of the element coefficients, expressed as rows in the $\mat A$-matrix.
This means that point and areal observations fit together in the same
framework \citep{moraga-etal-2017} without difficult covariance
calculations. This can be used, for example, to model both the
precipitation measurement at a weather station and river runoff from
catchment areas simultaneously. 
\red{It also simplifies the computation of excursion sets and uncertainty measures for contour maps, as shown by 
	\citet{bolin2015excursion} and \citet{bolin2017quantifying}.} 
The ease of
including areal observations means that the model in some cases can be
used instead of discretely indexed areal models 
(e.g. the Besag model) 
to provide a more realistic dependence
structure for a spatial model for regions of varying sizes or
spatio-temporal model for regions whose borders change at different
time steps during the period of interest.
The point process models in Section 5.3. are also based on areal observations.

\subsection{Parametrisations and priors}

The implementation in R-INLA needs interpretable parameters and good
default priors. The SPDE approach is most easily described through the
parameters $\tau$ and $\kappa$, where $\tau$ is the parameter for the
Mat\'ern covariance function that can be consistently estimated under
infill-asymptotics \citep{zhang2004inconsistent}, $\kappa$ shows up
naturally in the differential operator, and, under infill-asymptotics
the GRFs with parameters $(\kappa_0, \tau)$ and $(\kappa_1, \tau)$
have finite Kullback-Leibler divergence \citep{art590}. Internally in
R-INLA, computations are done using $\log(\tau)$ and $\log(\kappa)$
since these tend to give well behaved posteriors.

These parameters are unfortunately difficult to interpret, but the new
function \texttt{inla.spde2.pcmatern} uses marginal standard deviation
$\sigma$ and the empirical range $r = \sqrt{8\nu}/\kappa$ from
\cite{lindgren2011explicit} for R-INLA input and output. If we draw
multiple samples from the spatial component, $\sigma$ can be seen from
the variation in the spatial field and $r$ can be connected to the
typical distance between high and low regions. A joint principled
prior for $r$ and $\sigma$ was developed by \citet{art590} using the
penalised complexity framework developed by
\citet{simpson2017penalising}, and the prior has been successfully
applied in practice
\citep{beguin2017predicting,wakefield2017estimating}. The prior
shrinks towards a base model of infinite range and zero variance, and
includes two shrinkage rates that must be elicited from prior
information through specifying the tail probabilities
$\mathrm{P}(\sigma > \sigma_0) = \alpha_1$ and
$\mathrm{P}(r < r_0) = \alpha_2$.

\section{Spatial modelling with R-INLA}
\label{sect-spatial-modelling}

In the introduction we provided examples of a wide range of papers that
use R-INLA for model fitting. 
In this section, we discuss how combining the
computationally efficient representation of the spatial effect with
	general features of R-INLA makes fitting these many
different types of spatial models possible.

\subsection{Spatial GLMs and GLMMs}

The simplest non-trivial continuously indexed spatial model that can
	be fitted in R-INLA is the standard Gaussian geostatistical model in
Equation \ref{eq:simple-geostat}. 
There is a large literature on
different approaches for making this model computationally feasible
for large datasets and a review of them is given by
\citet{heaton2017methods}.
They find that all the methods perform well
and that the SPDE approach implemented in R-INLA performs best for the
chosen real-world dataset.
However, the main
advantage of the SPDE approach over the rest of these methods is that
it is implemented in R-INLA where complex spatial models are easy to create, 
within the latent Gaussian model framework.

The simplest extension of the Gaussian geostatistical model is to
create spatial GLMs by changing the observation process to
\begin{equation}
    y(\vec s_k) | \eta(\vec s_k), \theta \sim f(y(\vec s_k); \eta_k),
    \label{eq:spatial-cond-lik}
\end{equation}
where $\eta(\vec s_k)$ is the spatial signal
$X(\vec s_k) \beta + u(\vec s_k)$, and $f$ is the desired likelihood.
Spatial models for count data can be achieved through the Binomial,
Negative Binomial or Poisson likelihoods, and non-Gaussian observation
processes can be handled, for example, through t-distribution,
skew-Normal and Gamma likelihoods. R-INLA also supports simple
zero-inflated spatial models for count data
\citep{musenge2013bayesian,haas2011forest} or spatial hurdle models
for continuous responses
\citep{quiroz2015bayesian,sadykova2017bayesian}. Furthermore, spatial
generalised linear mixed models (GLMMs) are easily created by adding
other unstructured and structured random effects, and survival models
are supported through a parametrized likelihood such as the
Exponential, the Weibull, or a Cox proportional hazards model
\citep{martino2011approximate}.
All models are fast to compute, as the Laplace approximation, 
sparse matrix libraries and numerical optimisation routines,
enable us to avoid MCMC completely.

\subsection{Joint modelling}

\red{
Spatial GLMs and GLMMs can be made more complex by taking
advantage of the possibility in R-INLA of using different likelihoods
for different subsets of the observations. Divide the observations
$y(s_1), y(s_2), \ldots, y(s_N)$ into $G$ groups, where $g[k]$ denotes
group $k$, and assign likelihoods
$f_1, f_2, \ldots, f_G$ for groups $1, \ldots, G$, respectively.
Equation \eqref{eq:spatial-cond-lik} then generalises to
\begin{equation}
    y(\vec s_k) | \eta(\vec s_k), \theta \sim f_{g[k]}(y(\vec s_k); \eta_k)
\end{equation}
and allows us to create multivariate models for multiple responses
through a shared component structure
\citep{mathew2016reparametrization,ntirampeba2017joint}, to jointly
model the response and a misaligned spatial covariate
\citep{sadykova2017bayesian, barber-etal-2016}, to model a
spatial point pattern together with marks or covariates
\citep{illian2012toolbox,simpson2016going} and to model replicated point patterns \citep{art588}, see below.
}

\subsection{Spatial point processes}
\red{
Spatial point patterns are a third type of spatial data structure, which is different
from both areal and point-referenced observations. 
For point pattern
data, the locations of objects or events in space (the ``points'') are the observations of interest, 
and one typically aims to learn about the
mechanisms that generated the spatial pattern 
formed by the locations of the objects or events represented 
by these points \citep{moelleral:98, diggle:03, illian2008statistical}. 
With point-referenced data, the locations are considered fixed, 
but the values are considered random; 
for point processes, the locations are considered random,
and additional measurement on the objects or events may or may not be available.
These values are called marks in the point process 
terminology, and a point pattern with marks is 
called a marked point pattern.
Point processes may be characterised by a density 
function $\lambda(s)$, termed the intensity function, which 
we assume to be piecewise continuous, with 
$$\label{intfunc} \Lambda(B)=\int\limits_B\lambda(s){\rm d}x\,$$
for $s\in B \subset \mb R^d$, where 
$\Lambda$ is referred to as the intensity measure.
}

\red{
Different classes of point process models have been discussed in the literature.
These range from the simple homogeneous Poisson process, 
which represents uniform spatial randomness, 
to more complex models that generate aggregated 
patterns or patterns exhibiting repulsion among 
points \citep{lieshout:00}. 
The class of log-Gaussian Cox processes \citep{moelleral:98}, 
may be interpreted as latent Gaussian models (with a log link)
and hence may be fitted with R-INLA. 
These are doubly-stochastic inhomogeneous 
Poisson processes where the
log-intensity is a geostatistical model, 
suitable for modelling aggregated point patterns that result from 
observed or unobserved spatial covariates. 
For example, the
log-intensity may be described by an 
intercept, $\mu$, and a spatial
signal, $u(\vec s)$ of interest, e.g.
\[
    \log(\lambda(\vec s)) = \mu + u(\vec s).
\]
Then 
$\lambda(\vec s)$ gives an inhomogeneous Poisson process. Each
realisation of the process conditionally on $\lambda(\vec s)$ is a
different point pattern.
}

\red{
These models were originally available in R-INLA through discretely indexed models using a lattice
\citep{illian2012toolbox}. 
While this is a common approach in the literature, 
it is not using all of the information in the data as the point pattern 
is given as exact locations; the binning of points into grid cells has been shown as 
the major source of error in the lattice approximation \citep{simpson2016going}.
The continuously indexed formulation through the SPDE approach avoids these 
issues and yields all the advantages of continuously indexed models. 
The likelihood of a log-Gaussian Cox process is analytically intractable 
and needs to be approximated.  
The approach discussed in  \cite{simpson2016going} uses a  
numerical integration that is of Poisson form and hence 
tractable in R-INLA. 
However, this additional approximation makes specifying spatial point 
process models in  R-INLA more cumbersome than those for point-referenced and areal data. 
Applications of these models range from
presence-only data in ecology \citep{renner2015point} to modelling eye
fixations \citep{barthelme2013modeling}. 
}

\red{
Point process models are particularly relevant in ecology where 
there is a strong interest in understanding the spatial distribution 
and abundance of individuals or groups of individuals in space. 
However, in many cases the usual assumption of either ``window sampling''  
or the ``small world model'' \citep{baddeley2015spatial} does not necessarily hold. 
In particular in animals studies data are often collected along 
transects that cover only a very small subarea of the area of interest, 
and animals might not be detected uniformly across space. 
It is hence unlikely that the pattern has been fully observed, 
neither as small portion of an infinite pattern nor as a finite 
process that lives within a fixed and bounded region. 
Recent work has developed modelling approaches that 
account for complex observation processes, 
including varying detection probabilities  \citep{yuan2017point}.
}

\subsection{Space-time models}

The spatial effect can be made spatio-temporal by using the Kronecker
product models from Section \ref{sect-background}, resulting in a
separable covariance structure. The first application in R-INLA to
continuously indexed models was by \citet{cameletti2011comparing} who
used a simple geostatistical model with Gaussian responses where the
spatio-temporal interaction effect was described by combining the SPDE
approach for space with an autoregressive process of order 1 for time.
For high temporal resolutions, the approach can be computational
expensive, but a piece-wise linear approximation to the temporal part
of the spatio-temporal interaction can be employed to reduce
computational cost
\citep{book125,wakefield2017estimating}. Current
research towards non-separable models will be discussed in Section
\ref{sect-elias}.

\section{Adding complexity to the spatial effect}
\label{sect-advanced}

The Mat\'ern covariance structure is stationary and isotropic, which
means that for any pair of locations, the covariance is only dependent
on the distance between the locations. This is an idealization that
makes it easier to construct valid covariance functions, to
parametrise the covariance functions and to fit the resulting models.
However, stationarity and isotropy are strong assumptions and rarely
believed to be completely true, but, constructing the complex spatial
covariance functions required is challenging. Using non-Euclidean
spaces, and extending to spatio-temporal covariance functions adds
further complexity, because if even one pair of locations has a
covariance that is incompatible with the rest, the covariance function
is not valid. A key feature of the SPDE approach is that the
covariance structure is modelled through a differential operator and
the validity of the global covariance structure is ensured. In this
section we provide several examples of how more complex covariance
structure can be achieved for the spatial effect by simple local
changes to the differential operator of the SPDE, and we discuss how a
non-Gaussian dependence structure can be achieved for the spatial
field.

\subsection{Adding covariates in the covariance structure}

\begin{figure}
        \centering
        \includegraphics[width=.4\linewidth]
        {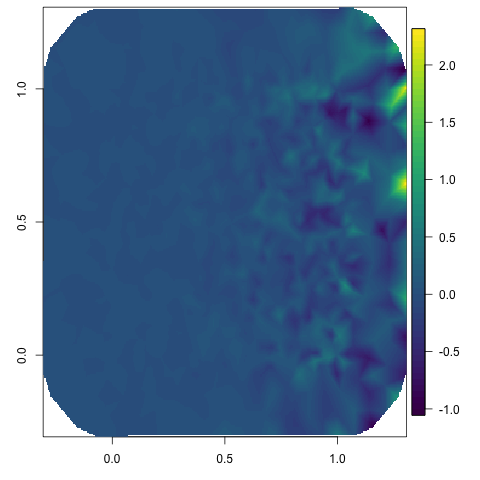}
        \includegraphics[width=.4\linewidth]
        {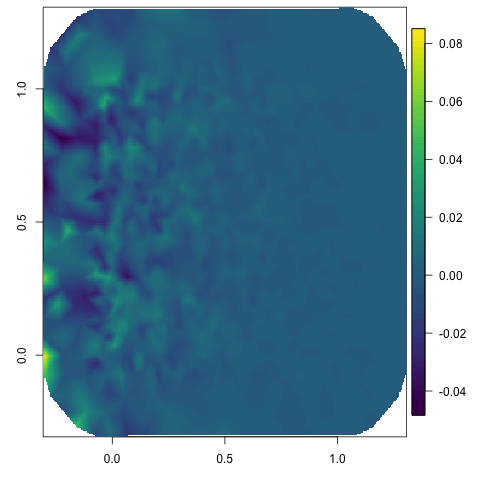}
        \caption{Simulation from the model by
            \cite{ingebrigtsen2014spatial}, as you decrease or
            increase $\tau$, increasing or decreasing the variance,
            from west to east. }
        \label{fig-ingebr}
\end{figure}

When modelling precipitation, we know that mountains influence the
spatial distribution of the amount of precipitation. In particular,
the amount of precipitation can be very different on either side of a
mountain due to, for example, orographic effects. To build more
realistic models for the covariance structure, we need to be able to
allow both the mean structure and the covariance structure to vary in
space. If covariates that can explain the variations are available, it
is useful to allow the covariance structure to vary as a function of
covariates, e.g.\ elevation. \citet{lindgren2011explicit} and
\citet{ingebrigtsen2014spatial} include the covariates in the
covariance structure by using the operator
$$\m L_2 = \left(\kappa(\vec s)^ 2 - \nabla\cdot\nabla \right)  \tau(\vec s),$$
where the parameters vary in space as sums of known basis functions $b_k(\vec s)$,
$$\log \tau(\vec s) = \theta_1^ \tau + \sum_{k=2}^{K} b_k(\vec s) \theta_k^\tau, $$
and similarly for $\kappa(\vec s)$.
See Figure \ref{fig-ingebr} for an example of a simulation from a model
with spatially varying marginal variance.

In practice, estimating both the mean structure and the covariance
structure from a single realization of a spatial process may lead to
inaccurate estimates and poorly identified parameters since the mean
structure and the covariance structure are not separately
identifiable. \citet{ingebrigtsen2015estimation} investigate
estimation of models with covariates in both the mean structure and
the covariance structure using multiple realizations. This model is
implemented in R-INLA with the options \texttt{B.tau} and
\texttt{B.kappa} in \texttt{inla.spde2.matern}.

\subsection{The Barrier model}

\begin{figure}
        \centering
        \includegraphics[width=.6\linewidth]
        {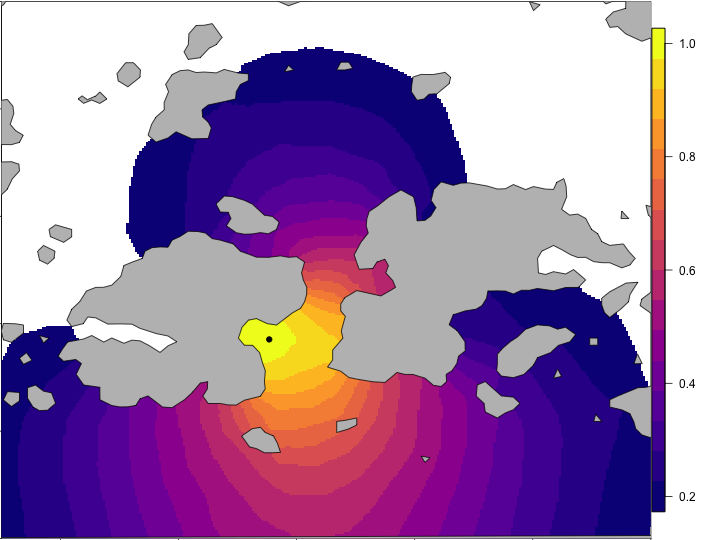}
        \caption{\red{Example correlation surface} for the Barrier model by \cite{bakka2018}.
                The grey region acts as a physical barrier to spatial correlation, forcing the model to smooth around this barrier.
        }
        \label{fig-barrier}
\end{figure}

When modelling aquatic animals near a coastline, the stationary model
smooths over islands and peninsulas, leading to unrealistic models.
Further, stopping the mesh at the coastline imposes the Neumann
boundary conditions, also leading to unrealistic models.
\cite{bakka2018} develop the Barrier model, defining the
operator
$$\m L_3 = \sigma^ {-1} \frac{2}{\pi r(\vec s)^ 2}
\left[ 1 - \nabla \cdot \frac{r(\vec s)^2}{8} \nabla \right], $$ with
$r(\vec s) \approx 0$ in a part of the study area, called the barrier
area, and $r(\vec s)= r$ in the rest of the study area. When modelling
aquatic animals, the land area is a physical barrier to spatial
correlation, see Figure \ref{fig-barrier} for an example simulation
where the GRF smooths around the barrier. Other applications may
include human activities on land, where water is a barrier, or they
may represent roads, power lines, residential areas, or ship traffic,
as physical barriers to a phenomenon. This model is implemented in
R-INLA as \texttt{inla.barrier.pcmatern}, and, since the sparsity of 
the precision matrix is the same as for the corresponding stationary model, 
the computational cost is roughly the same.

\subsection{Spatially varying anisotropy}

When modelling environmental data the assumption of isotropy may be
questionable because directional effects such as wind may cause higher
dependence in one direction than another. A simple way to achieve this
in a stationary model is geometric anisotropy. Geometric anisotropy is
equivalent to a linear transformation of the spatial coordinates and
can be achieved by replacing the differential operator with
$$\m L_4 = \kappa^ 2 - \nabla \cdot \mat H\nabla,$$
where $\mat H$ is a $2 \times 2$ positive-definite matrix.

\begin{figure}
    \centering
    \includegraphics[width=.8\linewidth]
    {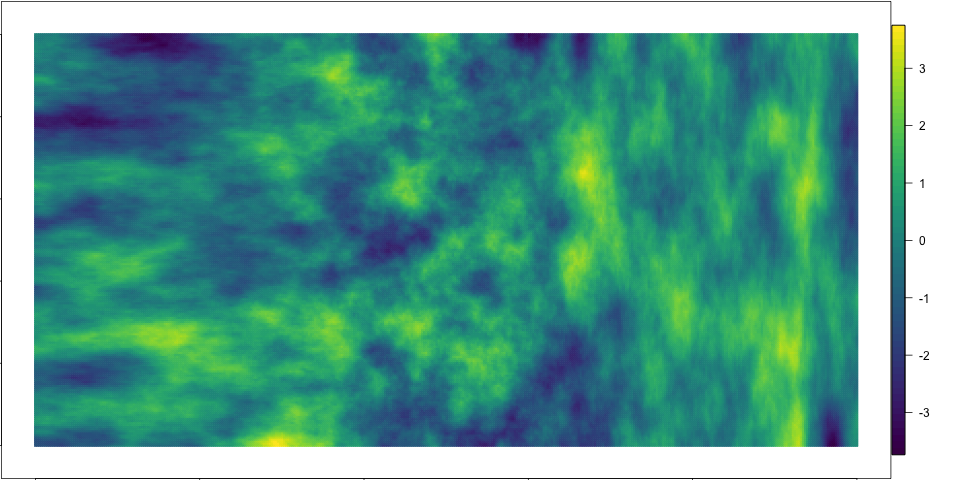}
    \caption{Simulation of anisotropic field from the model by
        \cite{fuglstad2015exploring}. In the west part of the plot
        there is a strong horisontal dependence, while in the east
        part there is a strong vertical dependence. }
    \label{fig-GA}
\end{figure}

However, directional effects such as wind may vary over the spatial
domain of interest and motivate non-stationarity in the anisotropy.
\cite{fuglstad2015exploring} discuss how to make the anisotropy
spatially varying by allowing $\mat H$ to vary spatially, and
\cite{fuglstad2015does} discuss a generalisation where both $\kappa$
and $\mat H$ vary and show that this controls both spatially varying
marginal standard deviations and spatially varying correlation
structure. Figure \ref{fig-GA} shows a simulation from a model where
the anisotropy varies continuously from extra horizontal dependence on
the left hand side to extra vertical dependence in the right hand
side. This model was not implemented in R-INLA,
when it was developed, but by using the new \texttt{rgeneric} 
framework \citep{art632} this is now possible.
SPDE models with locally
varying Laplacian can be interpreted as a change of metric in a
differentiable manifold, which leads to a substantial overlap with the
non-stationary models generated by the deformation method by
\citet{art244}.

\subsection{SPDEs on manifolds}

When modelling data on a global scale, using a rectangular subdomain
of $\mb R^ 2$ is problematic because it is hard to avoid singularities
near the poles and to construct covariance functions that make sense
when transformed to the true spherical geometry. Covariance functions
that are valid on $\mb R^3$ are valid on the sphere if chordal
distances are used, and some families of covariance functions are
valid when great circle distances are used, but validity of a
covariance function using great circle distance does not follow from
validity on $\mb R^2$ or $\mb R^3$ \citep{Huang2011}.

A major advantage of the SPDE approach is that if a two-dimensional
mesh can be created for the domain of interest, then the resulting
two-dimensional manifold structure can be used to interpret the
operator $\m L$ and the SPDE can be solved in the same way as for a
triangulation of a subdomain of $\mb R^2$. In theory, meshes can be
created for many kinds of two-dimensional manifold, but the important
one for handling global data is the sphere $\mb S^2$. This is
implemented as \texttt{inla.mesh.create(globe=...)} for a semi-regular
discretisation of the entire globe, but the same subdomain techniques
as on $\mathbb{R}^2$ can be used, after conversion of projected
coordinates, to Euclidean 3-dimensional coordinates. This conversion
can be automated by using the Coordinate Reference System (CRS)
specification that usually accompany large scale spatial data, that
specify which planar projection was used for the planar coordinates,
such as UTM or longitude and latitude. By specifying what coordinate
space the mesh should be built on, optionally with the aid of the
\texttt{inla.CRS} function, and providing data as spatial objects with
the \texttt{sp} package
\citep{pebesma2005sp,bivand2005appliedspatialwithR}, conversion
between projected and spherical coordinates is carried out behind the
scenes. The implemented SPDE models do not use the CRS information to
compensate for projection shape distortion, so the choice of space
determines a degree of local anisotropy. For large-scale phenomena on
the globe, building the analysis mesh as a subset of the Euclidean
sphere is therefore often beneficial, as it more closely resembles
reality. Any other manifold that is locally flat can in theory also be
used, but currently requires the user to supply their own
pre-generated mesh. It is also possible to generalise the operators
and the SPDE approach to higher dimensions, as gradients and
divergence have natural extension to 3 or more dimensions. In Figure
\ref{fig-sphere} we show an example simulation of the model in Section
\ref{sect-elias} on the sphere, and refer to
\citet{zhang2016seismicinversion} for an application to three
dimensional seismic inversion.

\subsection{Non-separable space-time model}
\label{sect-elias}

\begin{figure}
    \centering
    \includegraphics[trim={17cm 7cm 17cm 7cm},clip,
    width=.3\linewidth]
    {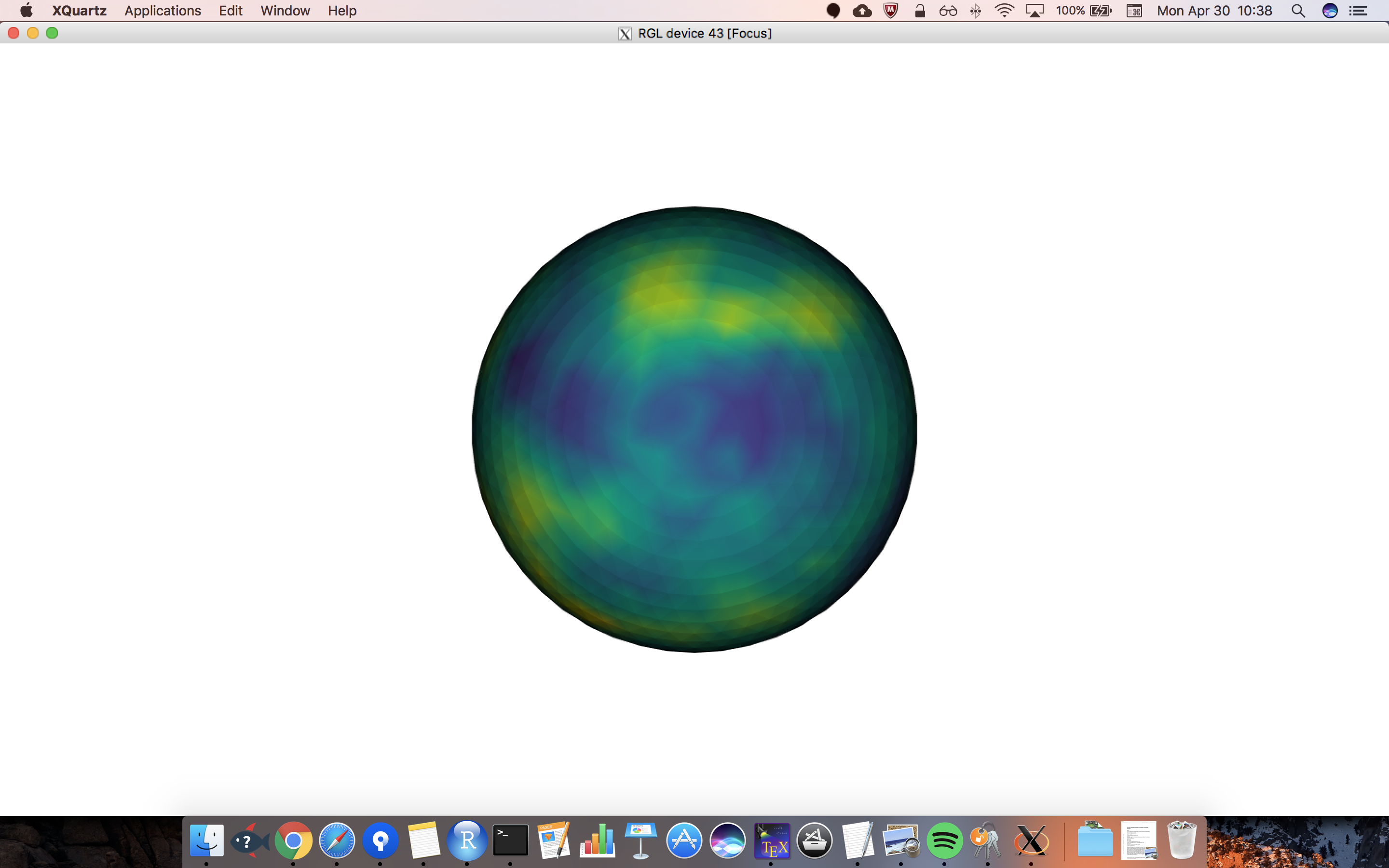}
    \includegraphics[trim={17cm 7cm 17cm 7cm},clip,
    width=.3\linewidth]
    {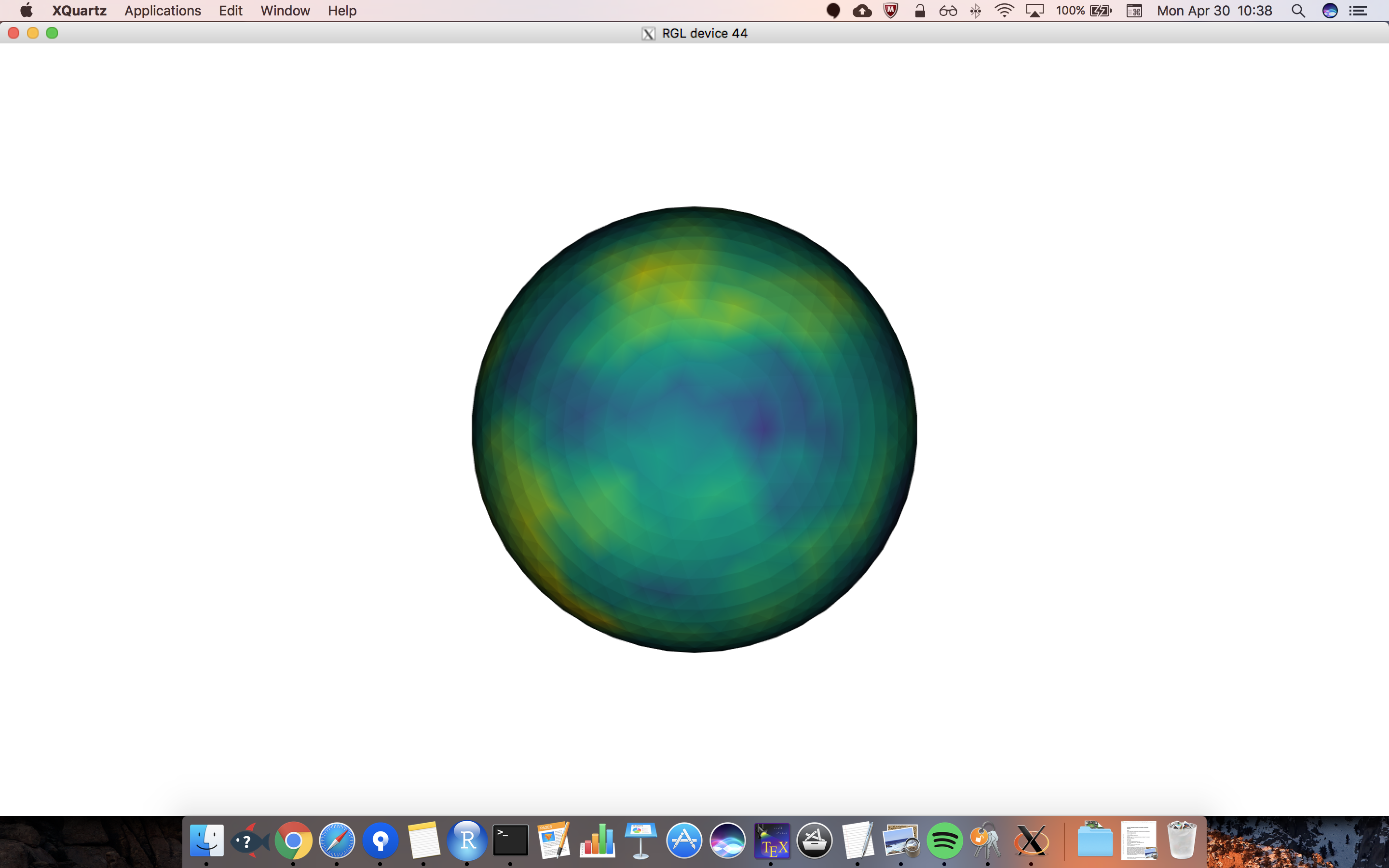}
    \includegraphics[trim={17cm 7cm 17cm 7cm},clip,
    width=.3\linewidth] {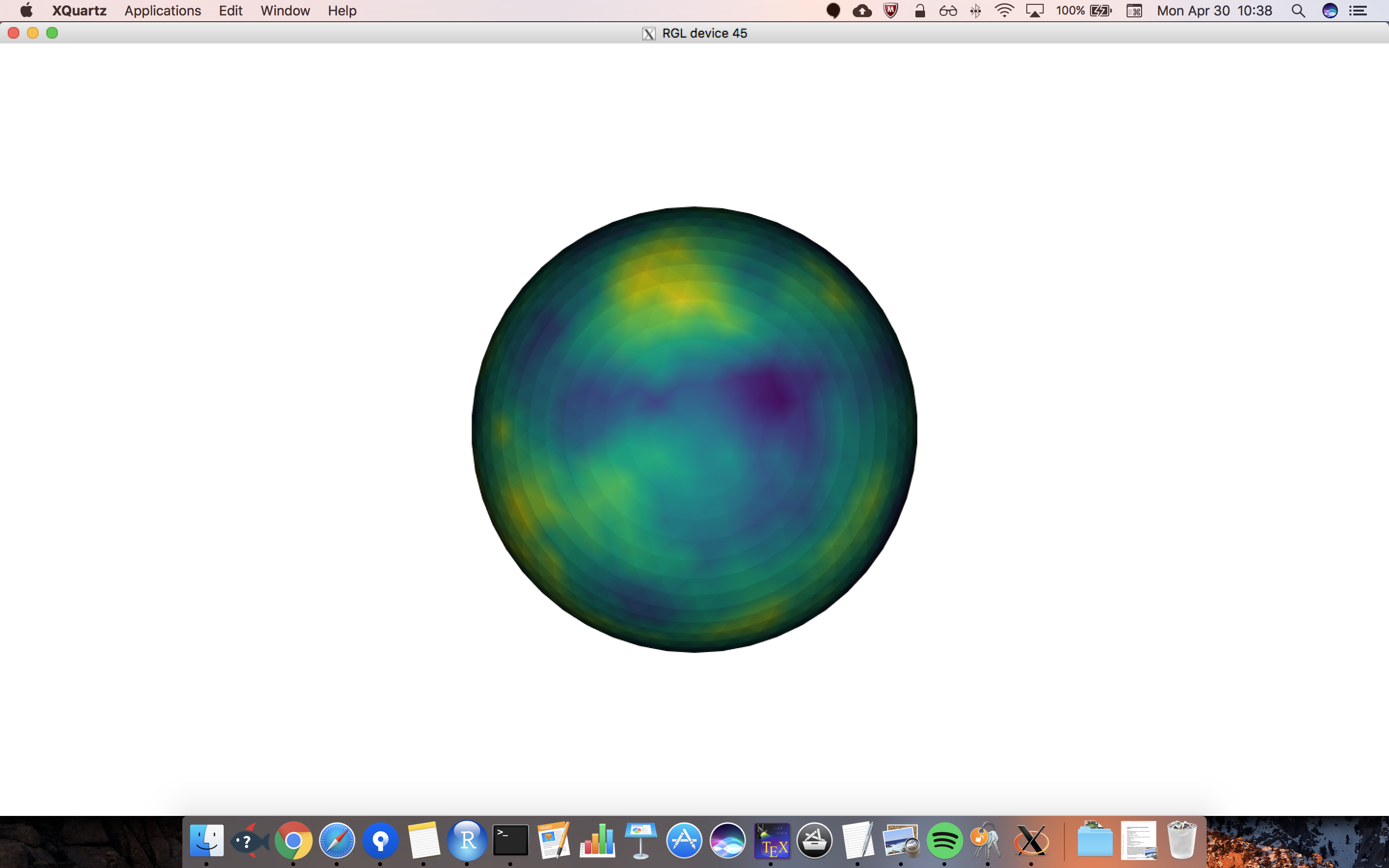}

    \includegraphics[trim={17cm 7cm 17cm 7cm},clip,
    width=.3\linewidth] {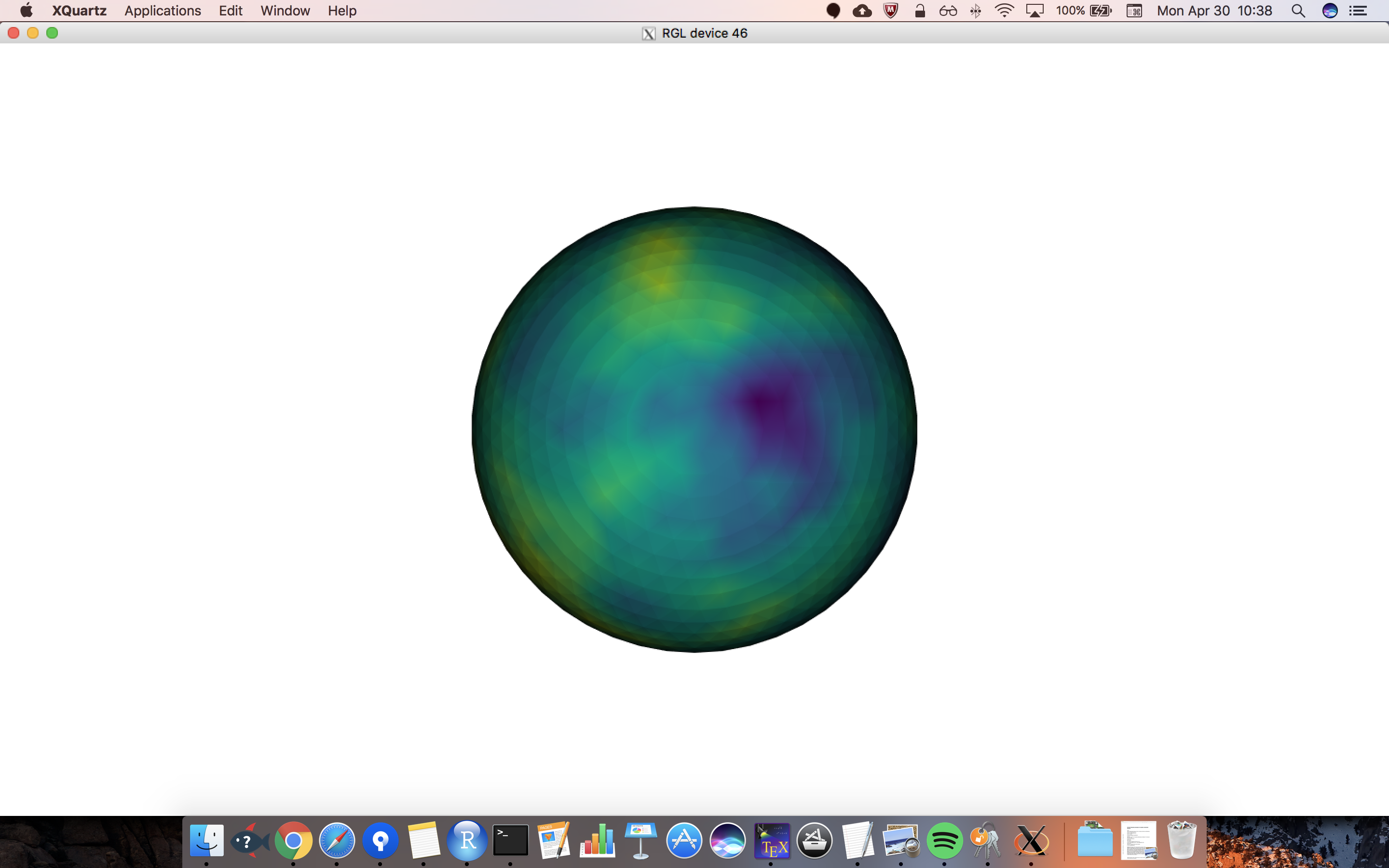}
    \includegraphics[trim={17cm 7cm 17cm 7cm},clip,
    width=.3\linewidth] {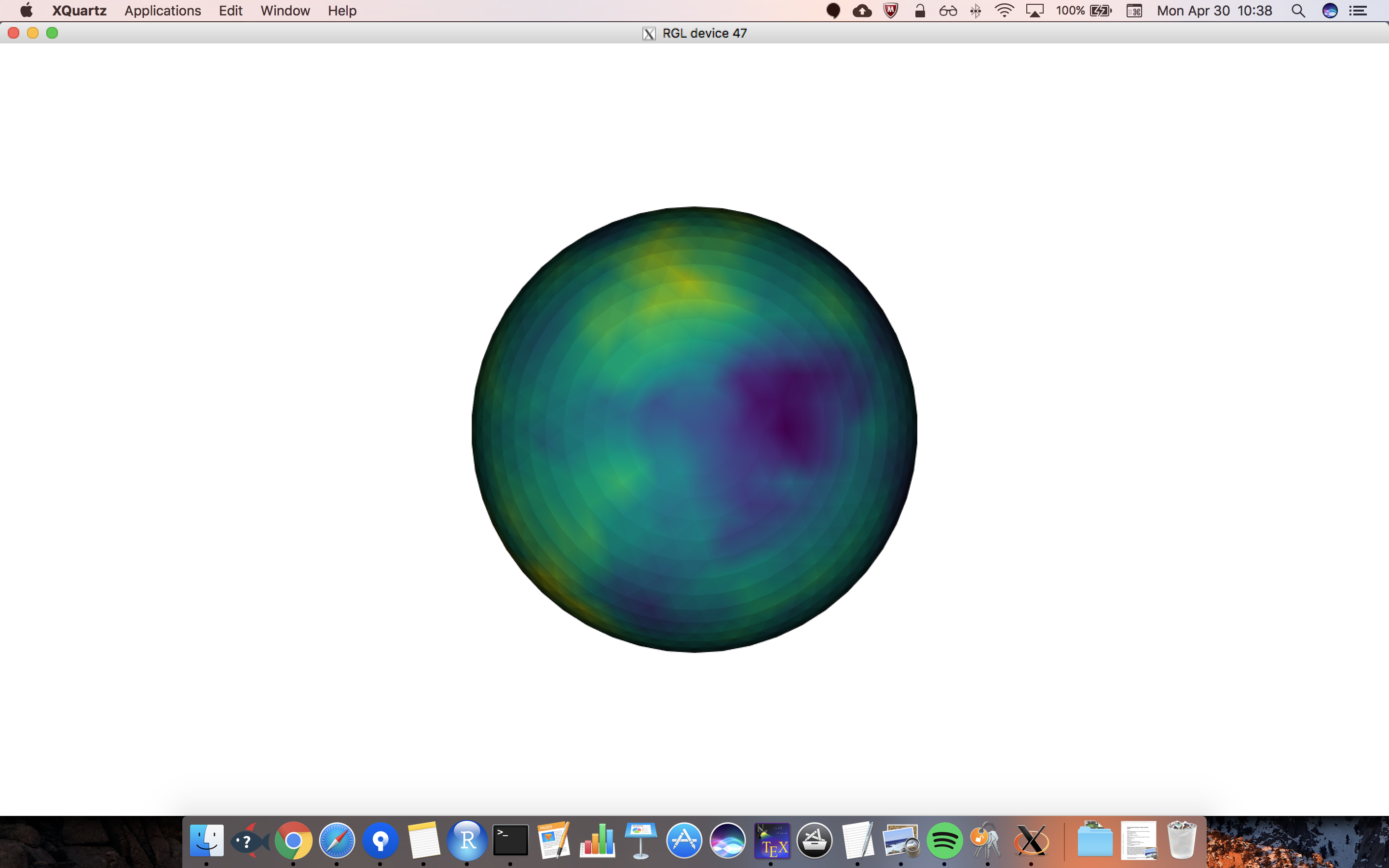}
    \includegraphics[trim={17cm 7cm 17cm 7cm},clip,
    width=.3\linewidth] {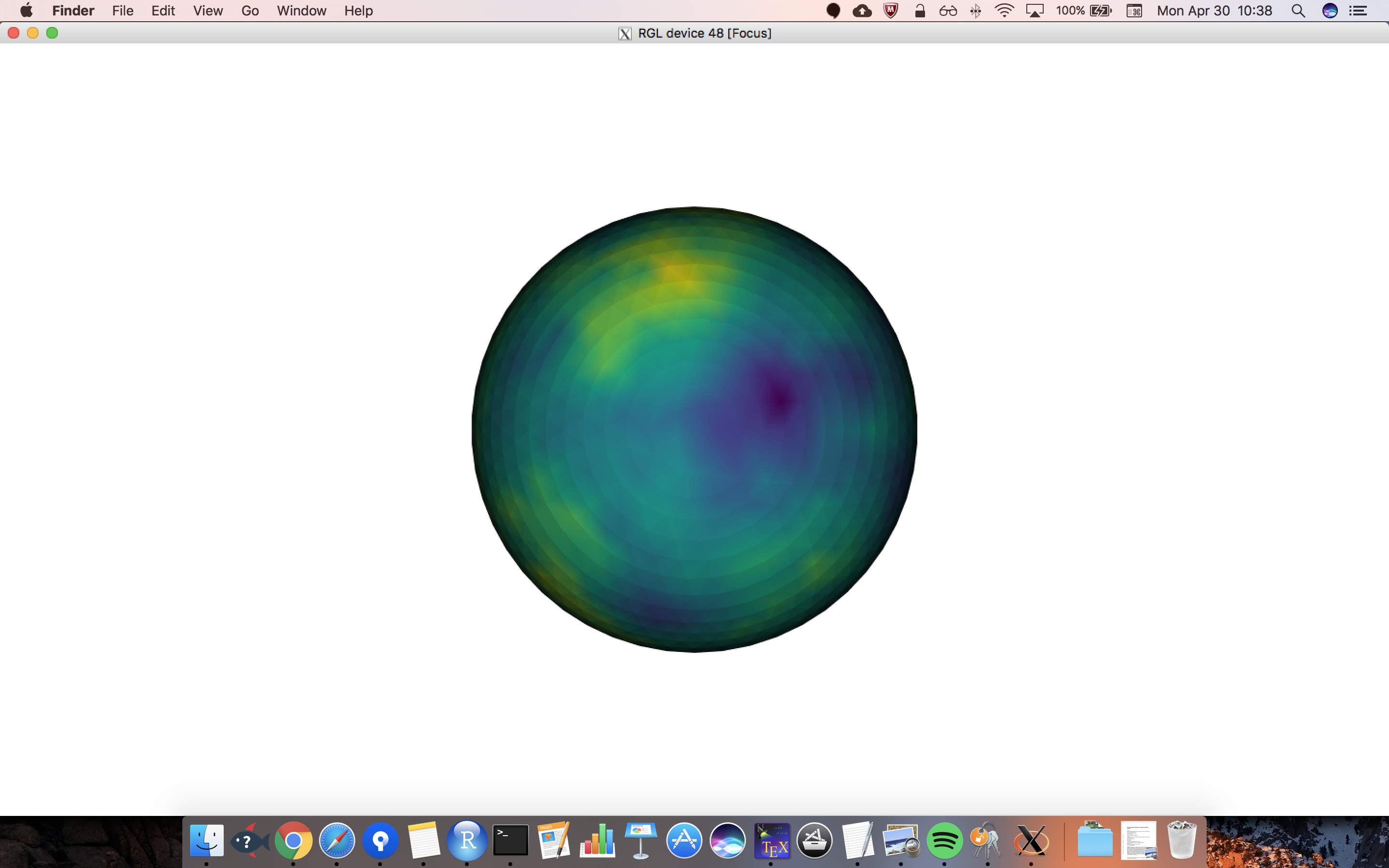}

    \caption{Simulation for the non-separable model by
        \citet{elias2018} on the sphere, for 6 time points (left to
        right, top to bottom). }
    \label{fig-sphere}
\end{figure}

Separable space-time models are convenient, but not always
appropriate; if we have a phenomenon that follows the heat equation,
the resulting field is non-separable. There are an infinite number of
ways non-separability can be described, but of special interest is a
non-separable model that is closely linked to the heat equation, as
this is one of the most common models in physics. \citet{elias2018}
refers to the separable space-time model that is the Kronecker of
Mat\' ern and AR(1) as
\begin{eqnarray}
  \m L_5 = 
  \left(\phi +\frac{\partial}{\partial t} \right)^{\alpha_t} 
  (1 - \gamma_{\mathcal{E}}\Delta)^{\alpha_{\mathcal{E}}/2},
\end{eqnarray}
and changes this to 
\begin{eqnarray}
  \m L_6 = \left(\gamma_t\frac{\partial}{\partial t}-\Delta\right)^{\alpha_t}
  (1 - \gamma_{\mathcal{E}}\Delta)^{\alpha_{\mathcal{E}}/2}
\end{eqnarray}
to produce a non-separable space-time model that is closely linked to
the heat equation. Applications of this include temperature modelling
on the globe, see Figure \ref{fig-sphere} for a simulated example. One
key advantage when considering a discretisation approach based in
\cite{lindgren2011explicit} is that the precision matrix for this
model has sparsity similar to the one for the separable model, thus
not adding computational burden. \citet{elias2018} consider some
marginal properties of the resulting multivariate Gaussian disribution 
to understand the model parameters.
Research is underway to understand for which applications this type of
non-separability is more appropriate than a separable model. This and
other related models are still development and we plan to implement
them in R-INLA in the near future.

\subsection{General smoothness}

The spatial operators we have discussed so far, $\mathcal{L}_i$,
$i=1,\ldots, 4$, are of second order. This explicitly determines the
differentiability of the resulting random fields $u(\vec s)$. To
define models with different smoothness, the operator can be replaced
by $\mathcal{L}_i^{\alpha/2}$, where $\alpha>0$ is a parameter
determining the smoothness of the field. In this case, a restriction
with the SPDE approach is that it is only computable if $\alpha$ is an
integer, as in equation \eqref{eq-matern-general}.
This is typically not a major restriction, since $\alpha$ is difficult to estimate, but may be important
in some cases \citep{stein2012interpolation}.

The parsimonious fractional approximation, which is implemented in R-INLA for the stationary Mat\'ern model, is not applicable for the more general non-stationary models. However, 
\citet{Bolin17rational} propose a rational SPDE method that is
computable for any $\alpha>0$, \red{and which has a higher accuracy than the parsimonious approximation for Mat\'ern model}. 
It combines the FEM approximation in
space with a rational approximation of the function $x^{-\alpha/2}$ in
order to compute an approximation of $u(\vec {s})$ on the form
$\vec{u} = \mat{P}\mat{x}$, where
$\vec {x} \sim \mathcal{N}(\mat {0},\mat {Q}^{-1})$, and $\mat {P}$ and $\mat {Q}$ are sparse matrices. 
This approximation facilitates including $\alpha$ as a parameter that is estimated from the data, and fits in the INLA framework, but is not yet included in the package.

\subsection{Non-Gaussian spatial fields}
\label{sect-non-gaussian}
\begin{figure}[t]
    \centering
    \includegraphics[width=0.6\linewidth]{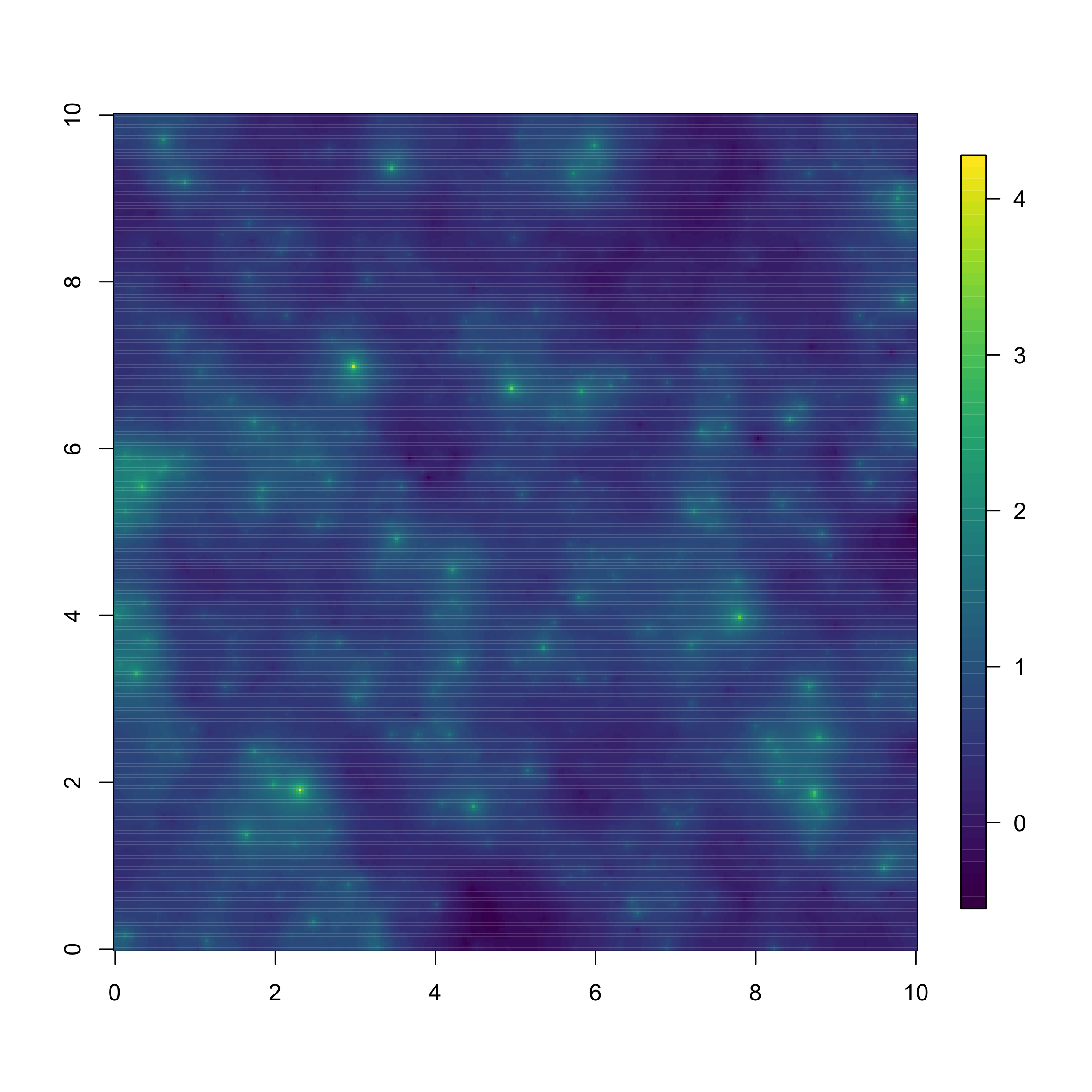}
    \caption{A simulation of a Mat\'ern SPDE model driven by NIG
        noise.}
        \label{fig-nig}
\end{figure}
If the process of interest has features that cannot be captured by a
Gaussian model, such as asymmetry in the sample paths or skewness in
the marginal distributions, non-Gaussian Mat\'ern-like fields can be
defined by replacing the driving noise $\mathcal{W}$ by other
non-Gaussian models. A simulation from a model like this can be seen
in Figure \ref{fig-nig}. The model that is used for the simulation is
$$
(\kappa^2-\Delta)(\tau u) = \mathcal{M},
$$
where $\mathcal{M}$ is normal inverse Gaussian (NIG) noise, see
\citet{Wallin15sjs} for a formal definition. As in the Gaussian case,
the process $u(\vec{s})$ has a Mat\'ern covariance function, but the
model has two additional parameters $\mu$ and $\gamma$ that
respectively control skewness and tails of the marginal distributions.
Using the parametrisation of the NIG noise from \citet{Bolin16mult},
to assure that $u(\vec{s})$ has zero mean, and discretising the model
using the FEM, yields the discretised model
\begin{equation}\label{eq:discretenig}
    \begin{split}
        \vec{u}|\tau,\kappa,\mu,\vec{v} &\sim
        \mathcal{N}(\tau^{-1}\mu\mat{\red{K}}^{-1}
        (\vec{v} - \vec{h}),\tau^{-2}\mat{\red{K}}^{-1}\diag(\vec{v})\mat{\red{K}}^{-T}),\\
        \vec{v}|\gamma &\sim IG(\gamma^2,\gamma^2 \vec{h}^2).
    \end{split}
\end{equation}
Here $\vec{v}$ is a vector of independent inverse Gaussian (IG)
distributed variables and $h_i = \db{\langle \phi_i, 1 \rangle}$.

Thus, the model is Gaussian conditionally on $\vec{v}$, where the
parameter $\mu$ controls the mean of $\vec{u}|\vec{v}$. Comparing
\eqref{eq:discretenig} with the corresponding Gaussian model
\eqref{eq-matern-general}, we note that the matrix $\mat{C}$, with
elements $C_{ii} = h_i$, has been replaced by a matrix with IG
distributed elements satisfying $\mathbb{E}(v_i) = h_i$. Because of
this representation, it is easy to simulate non-Gaussian SPDE models
using R-INLA. However, the INLA methodology cannot be used for
inference in this case since the models are intrinsically
non-Gaussian. Inference can instead be performed using stochastic
gradient methods \citep{Bolin16mult}.

\subsection{Additional work}

The models we have illustrated in this section are just scratching the
surface of the possibilities for extending the simple Gaussian and
isotropic covariance structures with the SPDE approach. Systems of
SPDEs have been used to create multivariate spatial fields
\citep{hu2016spatial,hu2013spatial,hu2013multivariate,Bolin16mult} and
to generate more flexible anisotropic and oscillating covariance
structures both on $\mathbb{R}^d$ and on the sphere
\citep{Bolin11nested, barman2017three}. \citet{yue2014bayesian}
develops adaptive Bayesian splines based on the SPDE approach.

\section{R-packages building on R-INLA}
\label{sect-cont-spatial-r-packages}

\red{
	R-INLA is freely available from \url{www.r-inla.org} and open source, and
	there are multiple R-packages for discretely indexed spatial models
	defined on top of R-INLA. Their goal is to make the models more easily
	accessible for the applied researcher and/or offer additional
	functionality. \citet{diseasemappingPackage, geostatsp} provide the
	package {\tt diseasemapping}, which allows the user to implement
	Poisson regression models incorporating fixed effects and either the
	BYM or BYM2 model including PC priors. The R-package {\tt SUMMER}
	\citep{SUMMER} uses the classical BYM model for small area estimation
	of under-5 mortality based on survey data, see also
	\citet{mercer-etal-2015}. The results of different spatial models
	fitted with R-INLA to the same dataset can be combined through the
	R-package {\tt INLABMA} \citep{art528}.
}

\red{
	Very recently, the Shiny app SSTCDapp
	(\url{http://www.unavarra.es/spatial-statistics-group/shiny-app}) has been
	introduced, which allows the estimation of different discrete space and
	space-time models using R-INLA in a user-friendly way without
	\texttt{R}-code \cite{ugarte-etal-2014, goicoa-etal-2017,
		adin-etal-2017}. The app provides different descriptive statistics
	and supports several spatial and temporal model components. 
	Furthermore, the four proposed space-time interaction
	types by \citet{knorrHeld-2005} are available, and the authors offer
	tutorials that explain the usage of the app.
}

\red{
Multiple packages for continuously indexed spatial models have also been
developed on top of R-INLA. 
\citet{geostatsp} provides an easier
interface to the SPDE models for geostatistical modelling in the
package \texttt{geostatsp}, and \citet{bolin2015excursion} provides a
package for calculating credible regions and excursion sets based on
the output from R-INLA in the package \texttt{excursions}
\citep{bolin2018excursions}. The \texttt{inlabru} package
\citep[][\url{https://inlabru.org/}]{bachl2017inlabru} provides a new
general R-INLA user interface that in particular simplifies the
specification of spatial and spatio-temporal models, by hiding all the
\texttt{inla.stack} code from the user. It also extends the available
class of models to mildly non-linear predictor models, and provides a
\texttt{predict} function for non-linear posterior prediction based on
i.i.d.\ posterior samples. Since it was initially developed for
ecological survey data, it has special support for point process data. 
All these packages
are available on \texttt{CRAN}.
}

\section{Discussion}
\label{sect-discussion}

Reformulating spatial models into a form that is suitable for R-INLA
can be challenging, but has great rewards. Within R-INLA, we can
easily combine the spatial effect with other random effects, fixed
effects or complex likelihoods to create complex spatial or
spatio-temporal models. The INLA method allows fast Bayesian inference
and makes full Bayesian inference possible for models that before were
considered infeasible, or would require time-consuming and careful
implementation of a sampler. Further, the speed-up for small models
means that they can be run multiple times, e.g.\ for cross-validation.
An important advantage for users is that adding model components,
using non-Gaussian observation likelihoods or multiple likelihoods,
and extending to separable space-time models, require little
additional implementation effort.

The INLA methodology is centered around precision matrices, and CAR
models like the Besag model, which by design have sparse precision
matrices, are automatically well-suited for R-INLA. Continuously
indexed models are more challenging, as they must be discretised in a
form that admits a sparse CAR structure and parametrising CAR models
directly is not a robust approach. However, the SPDE approach allows
us to create a CAR structure for a continuously indexed discrete
approximation of the spatial field by discretising an SPDE. This adds
difficulty in understanding and setting up the problem, but we believe
the comment made by \citet{lindgren2011explicit} is still valid:
\begin{quote}
    \emph{...the approach comes with an implementation and
        preprocessing cost for setting up the models, as it involves
        the SPDE, triangulations and GMRF representations, but we
        firmly believe that such costs are unavoidable when efficient
        computations are required.}
\end{quote}
Purely applied users have little need to understand the FEM or SPDEs,
beyond learning how to create a reasonable mesh and the connection
between the spatial resolution and meshes. For statistical
researchers, learning the SPDE approach requires a more significant
effort and a more detailed study of the theory of Gaussian random
fields, but it also produces generally straight-forward and stable
results. Statistical models following the SPDE approach tend to work
as intended, without many surprises and without the need to make
clever---and often hidden---choices that are specialised for
individual cases. The development of the advanced models in
Section~\ref{sect-advanced} was not \emph{easy}, but the approach was
\emph{straight-forward}, in the sense that the research resulted in
stable algorithms with behaviour ``as intended''. The trade-off we have
from using the SPDE approach in our research is that, instead of
studying different tools for each generalisation of the model, we can
study general tools used in numerics and physics and ways to
discretise a differential operator, and apply this to the SPDE
formulation.

The development of computationally tractable advanced models in space
and space-time is not a simple task. We are fortunate to have many
inquisitive and demanding users who challenge us to expand our
research to fill the gaps that are necessary for good statistical
problem solving. One of the hotly disputed topics is the question of
what priors to use, including what interpretable parametrisations to
define them on, and we consider the development of default priors,
that we can stand behind, to be one of the biggest advances in R-INLA.
There are other topics where there is still much work left to do,
specifically for stability/robustness/sensitivity and for model
comparison criteria. 

Non-stationary models are often applied to space-time data, as
replicates are needed to infer the non-stationarity, automatically
putting them in competition with non-separable models. At this time,
the focus is on implementing and documenting different non-stationary
and non-separable models in R-INLA. The aim is to lay the groundwork
for a future literature rich with examples and comparisons for a wide
range of applications of these advanced models. After gaining the
necessary practical experience with these methods, we can start
answering the question of ``when you should use what model'', e.g.\ in
what applications is a separable space-time model with one specific
type of non-stationary spatial structure the most relevant model.

The main computational challenge for the future is space-time models.
At the time of this review, R-INLA can deal with applications with a
hundred thousand dimensions in the space-time representation. However,
having 100 nodes in each spatial and temporal direction gives a
million dimensions, which is where R-INLA fails, and 100 nodes is a
decent, but not very fine, resolution. Parallel computing may be able
to overcome this obstacle, and we are currently investigating adequate
approximations and factorisation methods that run well in parallel.

To sum up, for a particular application the question is almost never
``what is the right thing to do'', but rather ``what is the best thing
to do that can be implemented and run in the limited time available''.
The tools presented in this review have been successful in answering
this question for many different applications, especially those that
require more advanced models, and we continue to develop these tools
to push the boundaries of applied statistical modelling.

\section{Acknowledgements}
\red{
The authors acknowledge comments from G.\ Konstantinoudis and from the referees.
}

\bibliography{local-hrue,local,local-ar,lindgren,local-ds,local-jbi}

\appendix

\section{Appendix}

\subsection{Details on the BYM2 model}
\label{app-besag}
\red{
For the BYM2 model (see Section \ref{sect-besag-pri}), we propose to use the penalised
complexity (PC) framework developed by \citet{simpson2017penalising} to define priors
for $\tau$ and $w$. This implies the use of an exponential prior with
parameter $\lambda$ for the standard deviation $1/\sqrt{\tau}$. The
parameter $\lambda$ can be elicited using the prior probability
statement $\pi((1/\sqrt{\tau}) > U) = \alpha$, which gives
$\lambda = -\log(\alpha)/U$. Since $\tau$ represents the marginal
precision, it can be directly related to the total residual relative
risk, for which an intuitive interpretation is available.
\citet{simpson2017penalising} give the rule of thumb that the marginal standard
deviation of $\mat u$ with $\mat Q = \mat I$, after the exponential
distribution for $1/\sqrt{\tau}$ is integrated out, is about $0.31U$
when $\alpha = 0.01$. Believing for example that the residual relative
risk lies with a probability of $0.99$ within the interval
$(0.4,2.5)$, this would lead to an approximate interval
$(-0.92, 0.92)$ on the linear predictor scale assuming a log link. The
marginal standard deviation of the spatial effect is assumed to be
$0.92/2.58 \approx 0.36$ which gives parameters $U = 1.16$ and
$\alpha = 1\%$ in the PC prior for $\tau$.
}

\red{
The PC prior for the weight $w$, 
interpreted as the proportion of the variability contributed by the structured component,
is not available in closed form
\citep{simpson2017penalising}, but a probability contrast Prob($w < U) = \alpha$ can
be used to specify its parameter. The use of PC priors allows the
flexible, in some cases overfitting, model \eqref{eq:bym2} to shrink
sequentially towards two base models. The prior first shrinks towards
a model without any spatial variation, i.e.\ setting $\tau$ to
infinity, then shrinking towards a model with only unstructured
variability, i.e.\ setting $w=0$. This desired behaviour has been
shown by \citet{art585} and is not achievable by other commonly used
priors. We note that \citet{art585} recommend the use of the BYM2
model over the Leroux model. 
}

\subsection{Details on the approximation of the Mat\' ern field on a grid}
\label{app-grid}

\red{
Assume that the Gaussian random field $u$ is observed on an $M \times N$ grid, indexed by row
$i$ and column $j$, where the distance between neighbouring
observation locations are $h$ both vertically and horizontally. We can
use central differences to construct a discrete representation of
$\m L_1$,
\begin{align}
\mat L_1 &= \kappa^ 2 \mat  I + \mat D \\
D_{a, b} &= 
\begin{cases}
4/h^2, & a=b, \\ 
-1/h^2, & a\sim b, \\
0, & \text{otherwise},
\end{cases}
\end{align}
where $\mat I$ is an $MN \times MN$ identity matrix,
$a = (i,j) \sim b = (i^*, j^*)$ if $a$ and $b$ are first-order
neighbours, and periodic boundary conditions are used to avoid
boundary corrections for $\mat D$.
}

\red{
To visualise discrete operators, we use computational stencils,
$$
\mat I \stencil
\begin{tabular}{|c|c|c|}
\hline  &   &  \\ 
\hline & 1 &   \\ 
\hline  &  &  \\ 
\hline 
\end{tabular} 
\quad
\mat D \stencil h^ {-2} \
\begin{tabular}{|c|c|c|}
\hline  & - 1 &  \\ 
\hline -1& 4 & -1  \\ 
\hline  &  -1 &  \\ 
\hline 
\end{tabular} 
\quad
\mat L_1 \stencil h^ {-2} \
\begin{tabular}{|c|c|c|}
\hline  &  -1 &  \\ 
\hline -1& $\kappa^ 2 h^ 2+4$ & -1  \\ 
\hline  &  -1 &  \\ 
\hline 
\end{tabular} 
$$
where the central value is the value on the diagonal of the matrix,
the value above the centre is the matrix element for the entry
representing a grid cell and its neighbour to the north, etc. The
constant $h^ {-2}$ outside of the computational stencil is a constant
multiplying the entire matrix. Note that stencils are often
rotationally symmetric, and they may use larger neighbourhoods.
}

\red{
Since the concept of discretised differential operator is of
fundamental importance to the SPDE approach, we now present a detailed
numerical example in $\mb R^ 1$.
}
\begin{exmp}
	\red{
		Let the differential operator be the second derivative,
	$$\m L_{1\up D} = \frac{\d}{\d x} \frac{\d}{\d x} . $$
	If the operator is applied to the function
	$f(x) = 5x^2 +\sin(15 x)$, a new function
	\begin{align*}
	g(x) = \m L_{1D}(f)(x) 
	&=   \frac{\d}{\d x} \frac{\d}{\d x} f(x)  \\
	&=   10 - 15^2 \sin(15 x),
	\end{align*}
	is produced and the result is just the second-order derivative of
	$f$.
}
	
	\red{
	If $u$ is discretised to a grid where the distance between
	neighbouring grid cells is $h$, the operator $\m L_{1\up D}$ can
	be discretised to a matrix $\mat L_{1\up D}$, with computational
	stencil
	$$
	\mat L_{1\up D} \stencil h^ {-2} \
	\begin{tabular}{|c|c|c|}
	\hline 1 & -2 & 1  \\ 
	\hline 
	\end{tabular} .
	$$
	The stencil corresponds to the standard central
	difference approximation to the second-order derivative,
	\begin{align*}
	g(x) &= \lim_{h \to 0} \frac{f'(x+h/2) - f'(x-h/2)}{h} \\
	&=\lim_{h \to 0} \frac{(f(x+h) - f(x)) - (f(x) - f(x-h))}{h^ 2} \\
	&\approx  h^ {-2}  (1f(x+h) - 2f(x) + 1f(x-h)).
	\end{align*}
	The stencil gives a matrix with $-2h^ {-2}$ on the diagonal,
	$1 h^ {-2}$ on the upper and lower diagonal, and 0 otherwise.
}

\red{
	A small example is shown in Figure \ref{fig-deriv}, where we have
	used an equally-spaced grid with $h = 0.1$, and compare $f(x)$ to
	$\vec f$, and $\mat L_{1\up D} \vec f$ to the values of
	$\m L_{1\up D} f$. The operator $\m L_{1\up D}$ was used by
	\citet{art435} to construct an approximation for the second-order
	random walk for irregular locations.
}

	\begin{figure}
		\centering
		\includegraphics[width=.4\linewidth]
		{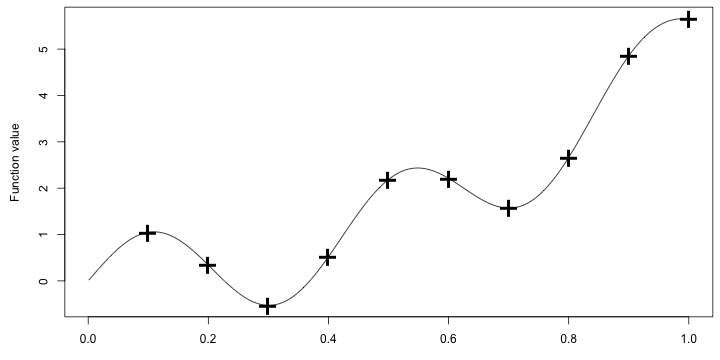}
		\includegraphics[width=.4\linewidth]
		{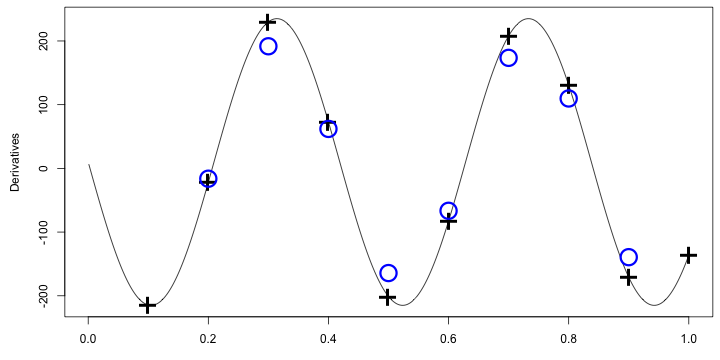}
		\caption{Example function $f$ (left plot) and
			$g=\m L_{1\up D} f$ (right plot). The black pluses are the
			vector $\vec f$ representing the continuous function $f$
			(black line), similarly with $\vec g$ and $g$. The blue
			circles is the discrete operator applied to the vector,
			i.e. $\mat L_{1\up D} \vec f$, which is close to the
			discretisation $\vec g$ of $g$. }
		\label{fig-deriv}
	\end{figure}
\end{exmp}

\end{document}